\documentclass[lettersize,journal]{IEEEtran}
\IEEEoverridecommandlockouts

\usepackage[utf8]{inputenc}
\usepackage[T1]{fontenc}
\usepackage{amsmath, amssymb, amsfonts} 
\usepackage{listings}                    
\usepackage{xcolor}                      
\usepackage{booktabs}                    
\usepackage{tabularx}                    
\usepackage{hyperref}                    
\usepackage[most]{tcolorbox}
\definecolor{codegreen}{rgb}{0,0.6,0}
\definecolor{codegray}{rgb}{0.5,0.5,0.5}
\definecolor{codepurple}{rgb}{0.58,0,0.82}
\definecolor{backcolour}{rgb}{0.95,0.95,0.92}

\lstdefinelanguage{Dafny}{
  keywords={method, function, returns, requires, ensures, invariant, decreases, if, then, else, var, while, set, int, bool, forall, exists, match, case},
  sensitive=true,
  comment=[l]{//},
  morecomment=[s]{/*}{*/},
  morestring=[b]",
}

\usepackage{cite}
\usepackage{enumitem}
\usepackage{amsmath,amssymb,amsfonts}
\usepackage{algorithmic}
\usepackage{graphicx}
\usepackage{textcomp}
\usepackage{comment}
\usepackage{booktabs}    
\usepackage{tabularx}    
\usepackage{caption}
\usepackage{xcolor}

\usepackage{tabularx}
\usepackage{multirow}
\usepackage{booktabs}

\usepackage[numbers,sort&compress]{natbib} 
           
\def\BibTeX{{\rm B\kern-.05em{\sc i\kern-.025em b}\kern-.08em
    T\kern-.1667em\lower.7ex\hbox{E}\kern-.125emX}}

\usepackage[acronym]{glossaries}
\makeglossaries

\newacronym{llm}{LLM}{Large Language Model}
\newacronym{llms}{LLMs}{Large Language Models}
\newacronym{rr2d}{RR2D}{RealRequirements2Dafny}
\newacronym{smt}{SMT}{Satisfiability Modulo Theory}
\newacronym{nl}{NL}{Natural Language}

\begin{document}

\title{From Natural Language to Verified Code: Toward AI Assisted Problem-to-Code Generation with Dafny-Based Formal Verification\\
{}
\thanks{}
}

\author{Md Erfan, Md Kamal Hossain Chowdhury, Ahmed Ryan, and Md Rayhanur Rahman\textsuperscript{$\dagger$}%
\thanks{Md Erfan, Ahmed Ryan, and Md Rayhanur Rahman are with the Department of Computer Science, The University of Alabama, Tuscaloosa, USA. Email: \{merfan, aryan9\}@crimson.ua.edu, mrahman87@ua.edu}%
\thanks{Md Kamal Hossain Chowdhury is with the Alabama Water Institute, The University of Alabama, Tuscaloosa, USA. Email: mhchowdhury@crimson.ua.edu}%
\thanks{\textsuperscript{$\dagger$}Corresponding author.}%
}

\maketitle

\begin{abstract}
    Large Language Models (LLMs) show promise in automated software engineering, yet their guarantee of correctness is frequently undermined by erroneous or hallucinated code. To enforce model honesty, formal verification requires LLMs to synthesize implementation logic alongside formal specifications that are subsequently proven correct by a mathematical verifier. However, the transition from informal natural language to precise formal specification remains an arduous task. Our work addresses this by providing the NaturalLanguage2VerifiedCode (NL2VC)-60 dataset: a collection of 60 complex algorithmic problems. We evaluate 11 randomly selected problem sets across seven open-weight LLMs using a tiered prompting strategy: contextless prompts, signature prompts providing structural anchors, and self-healing prompts utilizing iterative feedback from the Dafny verifier. To address vacuous verification, where models satisfy verifiers with trivial specifications, we integrate the uDebug platform to ensure functional validation. Our results show that while contextless prompting leads to near-universal failure, structural signatures and iterative self-healing facilitate a dramatic performance turnaround. Specifically, Gemma 4-31B achieved a 90.91\% verification success rate, while GPT-OSS 120B rose from zero to 81.82\% success with signature-guided feedback. These findings indicate that formal verification is now attainable for open-weight LLMs, which serve as effective apprentices for synthesizing complex annotations and facilitating high-assurance software development.
\end{abstract}


\section{INTRODUCTION}
Formal specifications are indispensable for rigorously defining program logic and facilitating automated reasoning about software correctness. Formal specifications transform ambiguous behaviors into precise semantics, creating a framework for quality assurance through procedure contracts, loop invariants, and assertions. As a result of this mathematical clarity, these specifications have become essential across diverse tasks ranging from software testing \cite{mesbah2011invariant} and model checking \cite{rushby1995model} to full-scale program verification.

Formal verification is increasingly adopted to develop high-assurance software by providing mathematical proof that programs strictly satisfy their specifications \cite{ter2024role} than traditional dynamic analysis methods such as testing or fuzzing \cite{dipu2024formalfuzzer}. Consequently, high-stakes domains, including security-sensitive infrastructure, cryptographic libraries, and autonomous aerospace systems, rely on these rigorous correctness guarantees to prevent critical vulnerabilities and operational disruptions \cite{paul2023formal}. Despite significant breakthroughs in Satisfiability Modulo Theory (SMT) solvers \cite{barrett2005smt}, writing program properties and proofs remains a creative, manual process requiring immense expertise. Developers must manually generate complex annotations, such as loop invariants and ranking functions, to enable automated verification tools to complete a proof. This manual process is tedious and time-consuming, often exceeding the effort required to write the executable code itself. For example, the verification of the seL4 microkernel project \cite{klein2009experience} required an eleven-person-year effort \cite{murray2013sel4}, while the verified code for CompCert C \cite{leroy2025compcert} is more than three times the size of the compiler itself \cite{leroy2009formal}. 

Concurrent with the evolution of formal methods, the rise of \gls{llm} based assistants has rapidly transformed modern software development workflows. AI-driven tools such as GitHub Copilot \cite{MicrosoftCopilot}, Cursor AI \cite{cursor2024}, and Amazon Q Developer \cite{amazonq2024} have accelerated programming tasks through natural language to code translation and intelligent autocompletion. By leveraging massive corpora of source code, these systems synthesize complex snippets from informal descriptions to automate traditional refactoring and implementation workflows. This shift has popularized vibe coding \cite{GoogleVibeCoding2024}, where developers rely on high-level intuition and conversational prompts to iterate on software rather than manual line-by-line implementation \cite{ray2025review}. However, the use of LLMs in software synthesis introduces a new set of reliability concerns because, despite their fluency, these systems frequently produce code that is syntactically plausible but semantically incorrect, a phenomenon known as hallucinations. Furthermore, LLMs often exhibit insufficient reasoning capabilities when dealing with complex algorithmic logic and remain susceptible to generating insecure patterns due to adversarial poisoning or inherent biases in their training data \cite{ji2023survey}. Consequently, the need to ensure the correctness and logical integrity of LLM-generated code has emerged as a fundamental challenge in the software engineering community. The problems necessitate a bridge between the synthesis of AI and the rigorous mathematical certainty of formal verification.

Furthermore, software requirements are typically written in natural language, which is often ambiguous and imprecise. Capturing complete specifications from such requirements is difficult, and there is currently a lack of direct mapping from natural language to formal specifications. Several formal verification languages exist in the literature, such as F*, Coq, Lean, and the Java Modeling Language (JML); however, we choose Dafny \cite{leino2010dafny} for our framework due to its unique balance of imperative programming and automated theorem proving. Dafny supports verification via \textit{Design by Contract} \cite{hoare1969axiomatic} using assertions, preconditions, and postconditions. However, even in Dafny, authoring formal specifications and auxiliary verification assertions remains difficult for developers \cite{noble2022more}. This challenge is exacerbated by a limited number of training data; while popular languages like Java and Python have over 5 million repositories on GitHub, the Dafny ecosystem has approximately 779 results \cite{github_dafny_2026}. This lack of large-scale data makes Dafny particularly challenging for LLM to synthesize correct code without producing hallucinations.

To bridge these gaps, we introduce \textbf{NL2VC-60}: \textbf{N}atural \textbf{L}anguage \textbf{2} \textbf{V}erified \textbf{C}ode dataset, a novel benchmark designed to evaluate the synthesis of formally verified code from complex, real-world requirements. We began by hand-authoring 60 high-quality Dafny programs to optimize our prompt generation strategies, specifically targeting the nuanced demands of competitive programming problems from the UVa Online Judge~\cite{uva_online_judge_2026}. The UVa Online Judge is an online automated judging system for programming problems, hosted by the University of Valladolid. Using this foundation, we evaluate randomly selected 11 distinct problem sets across seven leading open-source LLMs using three specialized prompting techniques. Existing work primarily focuses on small-scale, textbook-style problems supported by limited Dafny datasets and natural language inputs that rarely exceed 50 words. In contrast, our problem set overcomes these constraints by incorporating complex algorithmic challenges that require significantly more detailed specifications and extensive descriptive contexts. Our analysis of the resulting code led to the creation of the first comprehensive dataset of Dafny-specific compilation and verification errors in the literature, providing a unique resource for understanding model failure modes in formal methods. We are the first to integrate uDebug~\cite{udebug} community test suites to ensure rigorous functional correctness. uDebug is a community-driven platform designed for competitive programmers to validate their solutions against high-quality test suites. By combining community-driven testing with SMT-based formal proof, NL2VC-60 offers a new standard for code generation that balances complex natural language requirements with mathematical certainty.

\textit{The goal of this paper is to advance the frontier of AI-assisted NL problem-to-code generation by establishing a robust, Dafny-based formal verification framework that evaluates open-weight LLMs' ability to translate requirements into provably correct and functionally accurate software.}\\
\smallskip
\noindent The primary contributions of this work are as follows:
\begin{itemize}
    \item We conduct the first comprehensive empirical study of open-source LLMs synthesizing verifiable Dafny code from real-world requirements, evaluating seven \gls{llm} models across three prompting strategies to establish baseline performance in formal code generation.
    \item We introduce NL2VC-60 dataset, a novel benchmark consisting of 60 hand-authored programs to bridges the gap between simple textbook tasks and the nuanced demands of competitive programming tasks.
    \item We provide the first categorization of Dafny-specific compilation and verification errors in the literature, creating a diagnostic dataset of model failure modes to guide future improvements in the synthesis of formal verification.
    \item We establish a rigorous evaluation pipeline for functional correctness by being the first to integrate extensive uDebug community test suites, ensuring synthesized programs are both formally verified and correct across thousands of real-world edge cases.
\end{itemize}

Ultimately, the synthesis of verified methods remains a vast problem space, and this paper serves as an initial exploration of its potential. While our first two contributions address the end-to-end task of synthesizing code from narrative prompts, our third contribution, the systematic categorization of compilation and verification errors, highlights that \gls{llms} may be most effective when tackling specific sub-problems. These include generating formal specifications from natural language, synthesizing imperative code from existing contracts, or focusing exclusively on the annotation bottleneck by producing loop invariants and termination conditions. 

Our study results suggest that efforts in LLM-assisted coding should concentrate on generating verifiable programs, and that the combination of open-source models with formal verification techniques provides a cost-effective path toward high-assurance software development. Modern software development requires handling real-world requirements, yet existing literature focuses on small, textbook-style programming and algorithmic problems. However, our study relies on different algorithmic patterns, which fully reflect real-world software requirements. We can reduce this gap; by curating a more representative set of problems from the UVa Online Judge.  

The remainder of this paper is organized as follows. Section~\ref{sec:motivation} provides a motivational example. Section~\ref{sec:background} establishes the background on the Dafny language, open-weight LLMs, and the uDebug platform. Section ~\ref{sec:summary_literature} reviews relevant literature and existing benchmarks. Section~\ref{sec:approach} details our methodology, and Section~\ref{sec:results} presents our experimental results and a detailed analysis of model failure modes. Section~\ref{sec:findings_discussion} discusses the implications of our findings and potential threats to validity in Section~\ref{sec:threats_validity}. Finally, Section~\ref{sec:conclusion} concludes the paper.

\section{Motivational Example}
\label{sec:motivation}

We consider the \textit{Magic Formula} problem (UVa 11934) \cite{uva_11934} to illustrate the gap between traditional competitive programming and formal verification. The task requires counting how many values of a quadratic function $f(x) = ax^2 + bx + c$ are divisible by a divisor $d$ within the range $0 \le x \le L$. In a typical software engineering workflow, a developer might rely on sample test cases to verify their logic. However, such an approach is prone to off-by-one errors and boundary case failures that testing alone may not catch. By contrast, formal verification ensures that the counting logic remains correct across all possible integer inputs within the specified bounds, transforming a fragile test-based heuristic into a mathematically guaranteed solution.
\begin{lstlisting}[language=Dafny, caption={Dafny implementation (Magic Formula problem)}, label=lst:dafny_magic, basicstyle=\small\ttfamily, frame=single]
function Power(x:int, n:int): int
  requires n >= 0
  decreases n
{
  if n == 0 then 1 else x * Power(x, n-1)
}

method MagicFormula(a:int, b:int, c:int, d:int, l:int) returns (result:int)
  requires d > 0 && l >= 0
  // Formal Specification: The result must match the cardinality of the set
  ensures result == |set x:int | 0 <= x <= l && (a*Power(x,2) + b*x + c) % d == 0|
  // Boundary Case: Constant functions
  ensures (a == 0 && b == 0 && c % d == 0) ==> result == l + 1
{
    var x, count := 0, 0;
    while x <= l
      invariant 0 <= x <= l + 1
      decreases l - x
    {
        var value := (a*Power(x,2) + (b * x) + c);
        if value % d == 0 { count := count + 1; }
        x := x + 1;
    }
    result := count;
}
\end{lstlisting}
Our research proposes a shift from testing-based validation to formal synthesis. As shown in Listing~\ref{lst:dafny_magic}, the Dafny implementation goes beyond the imperative logic of the loop by defining a formal \textit{contract}. The \texttt{ensures} clause specifies the ground truth using mathematical set cardinality:
\begin{equation}
    result = |\{x \in \mathbb{Z} \mid 0 \le x \le L \land (ax^2 + bx + c) \equiv 0 \pmod d\}|
\end{equation}

This example motivates our work: by utilizing \gls{llms} to generate both imperative code and associated formal specifications, we can leverage SMT solvers to provide a mathematical guarantee of correctness. This approach effectively eliminates common algorithmic bugs that persist even after extensive testing on platforms like UVa Online Judge \cite{uva}.

\section{BACKGROUND}
\label{sec:background}
This section provides the theoretical and technical foundations for contextualizing our study of AI-assisted formal verification. We first discuss the unique architecture of the Dafny language and the inherent challenges of the specification burden. Then we discuss the current state of open-weight \gls{llms} and the iterative prompt engineering techniques used to optimize their reasoning. Finally, we introduce uDebug as a validation layer to ensure that formally verified programs remain functionally robust under real-world test cases.

\subsection{Dafny: A Verification-Aware Programming Language}
Dafny \cite{leino2010dafny, leino2012developing, dafny_official} is a verification-aware, statically typed programming language originally developed at Microsoft Research \cite{MSRDafnyProject} and currently supported by the Amazon Automated Reasoning group \cite{AmazonDafny2023}. Dafny bridges the gap between high-level programming paradigms, including imperative, functional, and object-oriented styles, and formal mathematical proof. A distinguishing feature of Dafny is its native support for \textit{Design by Contract}, employing Floyd-Hoare-style \cite{hoare1969axiomatic} verification using preconditions (\texttt{requires}), postconditions (\texttt{ensures}), and loop invariants.

To develop a verified program, developers provide formal specifications along with executable code. The Dafny static program verifier then checks the functional correctness of the implementation against these specifications. This is achieved by transforming the code into an intermediate verification representation (Boogie) \cite{le2011boogie}, encoding the conditions into predicate calculus, and invoking the Z3 SMT solver \cite{de2008z3} to prove their validity. In recent years, Dafny has been used by industry leaders like Amazon to verify AWS authorization logic \cite{cook2018formal} and by Intel for hardware encryption libraries \cite{wang2023codet5+}.

While Dafny ensures that the code does what the developer specifies, the difficulty lies in the specification burden. As illustrated in Listing~\ref{lst:dafny_magic}, a simple method to find the value of a quadratic function may require more lines of formal annotations (preconditions, postconditions, and invariants) than actual executable code. Researchers have observed that writing these auxiliary verification annotations remains the primary bottleneck in formal software development \cite{noble2022more}. If an LLM can successfully synthesize both the implementation and the proofs required for verification, the code generation could lower the barrier for high-assurance software engineering.

\subsection{Large Language Models and Open-Weight Models}
The landscape of Large Language Models (LLMs) has shifted from a dominance of proprietary APIs (like GPT-4 and PaLM-2) toward highly capable \textbf{Open-Weight Models} \cite{copet2025cwm}. Unlike closed models, open-weight models~\cite{copet2025cwm} such as \textbf{Llama 3.3}~\cite{grattafiori2024llama}, \textbf{Qwen 3}~\cite{yang2025qwen3}, \textbf{Gemma 3}~\cite{kamath2025gemma}, and \textbf{Gemma 4}~\cite{manik2026gemma} allow researchers to host the models locally, providing full control over parameters, token limits, and data privacy.

Recent advancements in these models have demonstrated that smaller, specialized architectures (e.g., \textbf{Qwen 3 Coder}~\cite{cao2026qwen3}) can rival proprietary models in code generation and logical reasoning tasks. However, applying these open-weight models to verification-aware languages such as Dafny remains an underexplored frontier. Because Dafny code is scarce in public training datasets compared to Python or Java, our research explores whether the reasoning capabilities inherent in modern open-weight architectures can generalize to the syntactic and logical requirements of formal verification.

\subsection{Prompt Engineering and Self-Healing}
Prompt engineering \cite{white2023prompt, giray2023prompt} is the systematic process of crafting inputs to align an \gls{llms} output with a specific technical task \cite{reynolds2021prompt}. In the context of Dafny, prompts must be engineered to be unambiguous and structured. We employ a tiered approach: starting from \textit{Contextless Prompting} to establish a baseline, moving to \textit{Signature Prompting} to provide structural anchors, and finally utilizing \textit{Self-Healing Prompting}. Self-healing~\cite{tihanyi2025new} mimics the human developer's workflow by feeding the Dafny verifier's error messages back into the LLM, allowing the model to iteratively repair its logic and specifications until verification is achieved.

\subsection{uDebug: Beyond Vacuous Verification}
A critical challenge in LLM-driven formal synthesis is vacuous verification, where a model generates weak or trivial specifications that pass the Dafny verifier but do not actually solve the intended problem. We integrate \textbf{uDebug}~\cite{udebug} into our evaluation pipeline to mitigate the problem. uDebug is a community-driven platform designed for competitive programmers to validate their solutions against high-quality test suites. By providing an accepted output for given inputs, uDebug allows us to perform a dual-layer validation:

\begin{itemize}
    \item \textbf{Formal Layer:} The Dafny verifier proves that the code is logically consistent with its formal specifications.
    \item \textbf{Functional Layer:} uDebug ensures the code is semantically correct by testing the generated code against extreme edge cases and boundary conditions contributed by the competitive programming community.
\end{itemize}

As noted by Professor Miguel Angel Revilla (Creator of UVa Online Judge), uDebug is a perfect complement for identifying critical inputs that break solutions~\cite{udebug}. In this research, we use uDebug to confirm that our synthesized, verified Dafny programs are also functionally robust in real-world scenarios type requirements.

\section{Summary of the Literature}
\label{sec:summary_literature}
This section contextualizes our research within the broader landscape of automated software engineering and formal verification. We review existing efforts in program synthesis, evaluate the evolving role of LLMs in formal methods, and compare current Dafny-centric benchmarks.
\subsection{Program Synthesis and Verification with Dafny}
In the last two decades, formal methods for software synthesis \cite{gulwani2017program} and verification \cite{ringer2019qed} have transitioned from esoteric research topics to practical industrial tools \cite{leino2010dafny, jones2021theories}. Modern tools like Dafny, SAW, and SPIN are now mature enough to support critical applications in encryption algorithms \cite{yang2023towards}, Ethereum Virtual Machine (EVM) bytecodes \cite{cassez2023formal}, scientific software, and quantum circuitry \cite{li2022qafny}. 

Despite this maturity, a barrier to adoption remains the scarcity of engineers trained in formal specification \cite{garavel20202020}. The gap has spurred research into automated support for Dafny, such as XDsmith for differential testing \cite{irfan2022testing} and techniques for generating counterexamples when verifiers fail \cite{chakarov2022better}. Our work builds on this momentum by exploring how \gls{llms} can bridge the gap between \gls{nl} requirements and verified Dafny code.

\subsection{LLMs for Formal Methods and Software Engineering}
\gls{llms} have been increasingly applied to automated proof synthesis and theorem proving, with models like Baldur \cite{first2023baldur} and Thor \cite{jiang2022thor} outperforming traditional heuristic tools like CoqHammer \cite{czajka2018concrete}. Beyond proofs, researchers have utilized LLMs to translate natural language into Isabelle/HOL \cite{wu2022autoformalization} and event graphs \cite{madaan2022language}. While general-purpose models like GPT-4 often struggle with algorithmic reasoning \cite{frieder2023mathematical}, specialized models such as Minerva \cite{narkawicz2017minerva, lewis2001model} demonstrate that domain-specific pre-training can mitigate these limitations. 

In the broader software engineering context, \gls{llms} now support code completion, repair, and test generation \cite{nashid2023retrieval, tufano2023automating}. Our research follows the philosophy of letting \gls{llms} generate plausible candidates while leveraging the Dafny verifier to guarantee correctness, effectively filtering out the hallucinations common in LLM-generated code.

\subsection{Benchmarking Dafny Generation}
Existing literature on LLM-based Dafny generation remains relatively limited in both dataset scale, limited word length for problems, and problem diversity. Prior studies have primarily evaluated models ranging from GPT-3.5 to the recent Llama 3.3 \cite{grattafiori2024llama}, Qwen 3 \cite{yang2025qwen3}, and Gemma 3 \cite{sellergren2025medgemma} on a narrow set of benchmarks. Table \ref{tab:literature_review} provides a comparative overview with limitations of existing literature.

\begin{table*}[t]
\centering
\caption{Comparison of Dafny Verification Datasets and Benchmarks}
\label{tab:literature_review}
\small
\begin{tabular*}{\textwidth}{@{\extracolsep{\fill}}l l l c l l p{4.5cm}}
\hline
\textbf{Work / Dataset} & \textbf{Input Type} & \textbf{Problem Type} & \textbf{Dafny} & \textbf{Size} & \textbf{Avg Code} & \textbf{Limitations} \\
\hline
Clover \cite{sun2024clover} & Short NL / Ann. & Textbook & Yes & 63--66 & $\sim$19 LoC & Simple problems, small scale \\
MBPP-Dafny \cite{austin2021program} & NL (short) & Basic Python & Yes & 164 & $\sim$19 LoC & Entry-level tasks only \\
HumanEval-Dafny \cite{banerjee2026dafnypro} & NL (short) & Algorithmic & Yes & 132 & $\sim$50 LoC & Still benchmark-style \\
DafnyBench \cite{loughridge2024dafnybench} & Mixed & Real + Textbook & Yes & 782 & $\sim$53 LoC & Limited human-written verified programs \\
TacoDafny \cite{wangtoward} & NL (Gen.) & Synthetic & Yes & Auto. & Varies & Synthetic, not real-world \\
ATLAS \cite{baksys2025atlas} & Alg. + Ref. & Algorithmic & Yes & Large & Varies & No direct NL to Dafny \\
SpecGen \cite{ma2025specgen} & NL & LeetCode & No & N/A & N/A & Uses OpenJML, not Dafny \\
\hline
\end{tabular*}
\end{table*}

As shown in the table, benchmarks like Clover \cite{sun2024clover} and MBPP-Dafny \cite{austin2021program} focus on textbook-level tasks with average lengths of only on average 19 lines of code. DafnyBench \cite{loughridge2024dafnybench} represents a greater effort with 782 samples, yet the research relies heavily on converted rather than native requirements. Besides that, existing work primarily focuses on small-scale, textbook-style problems supported by limited Dafny datasets and natural language inputs that rarely exceed 50 words. In contrast, our problem set overcomes these constraints by incorporating complex algorithmic challenges that require significantly more detailed specifications and extensive descriptive contexts. While frameworks like ATLAS \cite{baksys2025atlas} synthesize verified code using algorithmic references and test cases, our approach targets the direct synthesis of Dafny programs from \gls{nl} descriptions, addressing the complexity of requirements.

\section{Approach}
\label{sec:approach}
This study aims to bridge the gap between textbook-style benchmarks and real-world software requirements by evaluating \gls{llm} performance on complex algorithmic tasks curated from the UVa Online Judge. We focus on the transition from \gls{nl} requirements to formally verified Dafny code through tiered prompting and iterative repair.

\subsection{Research Questions}
We investigate the capability of LLMs to synthesize formally verified Dafny methods. We employ a tiered evaluation strategy to isolate the impact of structural hints and iterative feedback. We address the following research questions:

\begin{description}[style=multiline, font=\bfseries]
    \item[RQ1] - \textbf{(Contextless Prompting):} How effective are LLMs at synthesizing fully verified Dafny methods when provided only with a natural language description, without any formal structural hints?
    
    \item[RQ2] - \textbf{(Signature Prompting):} How does the provision of a formal method signature and accompanying functional tests affect the initial synthesis success rate compared to contextless prompting?
    
    \item[RQ3] - \textbf{(Self-Healing Capabilities):} To what extent can an iterative feedback loop recover failed synthesis attempts under varying initial conditions?
    \begin{itemize}[label=, leftmargin=0.5cm]
        \item \textbf{RQ3a (Self-Healing with Contextless Prompting):} Can LLMs repair verification failures when the initial attempt was generated from \gls{nl} alone?
        
        \item \textbf{RQ3b (Self-Healing with Method Signature):} Does the presence of a pre-defined method signature provide a superior result for the self-healing process, leading to higher repair rates than contextless healing?
    \end{itemize}
    \item[RQ4] - \textbf{(Error Analysis ):} To what extent can error descriptions help to overcome errors by using the signature prompt and the self-healing method? 
\end{description}

\subsection{Problem Curation and Abstraction}
This subsection details the systematic process of converting competitive programming tasks into standardized formal requirements to ensure high-fidelity evaluation of \gls{llm}.

\subsubsection{Test Dataset} 
We conducted our study using a collection of problems with rich natural-language descriptions and corresponding formally verified Dafny code. Existing literature often relies on small-scale, textbook-style datasets like MBPP or HumanEval; however, these focus on basic programming tasks with short specifications. To evaluate LLMs on complex, real-world requirements, we curated 60 set of problems from the \textbf{UVa Online Judge}, an automated judging system with thousands of competitive programming problems~\cite{uva}.
\subsubsection{Problems Generalization}
Unlike benchmarks that use one-line descriptions, UVa problems provide paragraph-level specifications involving different programming tasks, with average word counts exceeding 179. However, these problems are typically encumbered by what we term \textit{presentation flavor} details designed for contest environments (e.g., input formatting, multiple test case counts like $|t| < 15$, and arbitrary constraints like $|a| < 10^9$) that do not contribute to the semantic understanding of the requirement.

\begin{table*}[t]
\centering
\caption{Comparison Between Original UVa Problem Description and Generalized Generic Description}
\label{tab:problem_generalization}
\small
\begin{tabularx}{\textwidth}{l X X}
\toprule
\textbf{Component} & \textbf{Original UVa Problem Description (Competitive Flavor)} & \textbf{Generic Description (Requirement Focused)} \\
\midrule
\textbf{Description} & Some operators checks about the relationship between two values and these operators are called relational operators. Given two numerical values your job is just to find out the relationship between them that is (i) First one is greater than the second (ii) First one is less than the second or (iii) First and second one is equal. & Some operators checks about the relationship between two values and these operators are called relational operators. Given two numerical values your job is just to find out the relationship between them that is (i) First one is greater than the second (ii) First one is less than the second or (iii) First and second one is equal. \\
\addlinespace
\textbf{Input} & \textcolor{blue}{First line of the input file is an integer $t$ ($t < 15$) which denotes how many sets of inputs are there. Each of the next $t$ lines} contain two integers $a$ and $b$ \textcolor{blue}{($|a|, |b| < 1000000001$)}. & The input contain two integers $a$ and $b$. \\
\addlinespace
\textbf{Output} & \textcolor{blue}{For each line of input produce one line of output. This line} contains any one of the relational operators '$>$', '$<$' or '$=$', which indicates the relation that is appropriate for the given two numbers. & The output contains any one of the relational operators '$>$', '$<$' or '$=$', which indicates the relation that is appropriate for the given two numbers. \\
\addlinespace
\textbf{Sample Input} & \textcolor{red}{3} \newline 10 20 \newline 20 10 \newline 10 10 & 10 20 \newline 20 10 \newline 10 10 \\
\addlinespace
\textbf{Sample Output} & $<$ \newline $>$ \newline = & $<$ \newline $>$ \newline = \\
\bottomrule
\end{tabularx}
\end{table*}

To adapt these for formal verification, we performed a \textbf{Generalization Process} in Table~\ref{tab:problem_generalization} to make the problems generic. We manually transformed the competitive programming descriptions into a more generic, requirement-focused form by removing presentation-specific instructions while preserving the core computational logic. For example, a problem asking to identify relational operators (``$>$'', ``$<$'', or ``$=$'') between two integers was stripped of its ``process $T$ lines'' loop instructions and reduced to its functional essence.

\subsubsection{Empirical Problem Selection} 
We selected 60 problems based on their acceptance rates and user submission statistics to act as a proxy for practical relevance. Across the selected problems, total submissions varied from approximately 27,000 to over 370,000 in the online Judge platform, ensuring the tasks were neither trivial nor excessively niche. 

For each generalized problem, we developed ground-truth Dafny verification code by all the authors to ensure a unified and satisfactory implementation standard. We utilized the \textbf{uDebug} platform to access a diverse collection of test cases, including edge and corner cases, to validate the functional correctness of our verified solutions. This resulted in our main test dataset, NL2VC-60: a collection of 60 real-world algorithmic problems, each consisting of a generic requirement description, a formal method signature, and a suite of validation test cases. Since no public Dafny implementations of UVa problems existed prior to this study, our dataset minimizes the risk of data leakage during LLM evaluation. 

\subsection{Human Written Dataset: NL2VC-60}

To perform \textit{Dynamic Few-Shot Prompting}, we required a high-quality, diverse collection of verified Dafny methods to serve as in-context exemplars. Given the absence of existing datasets that map complex natural language requirements to Dafny, we manually developed a reference set, \textbf{NL2VC-60}, consisting of 60 problems from our suite.

This process involved translating the core computational requirements of 60 UVa problems into complete Dafny implementations. Unlike standard coding tasks, this required the manual formulation of formal specifications, including method preconditions (\texttt{requires}), postconditions (\texttt{ensures}), and complex loop invariants until the Dafny verifier could formally prove the code's correctness. We performed all necessary annotations, hint insertions, and structural refinements until the verifier signaled a successful proof for each method.

In developing this dataset, we experienced first-hand the significant cognitive load associated with formal specification. Formulating precise postconditions that capture the semantic intent of paragraph-level requirements and providing sufficient invariants for algorithmic logic proved to be a rigorous undertaking. The 
creation took approximately 300 person-hours for the authors and 50 more hours to resolve conflicts among authors to create this dataset of 60 formally verified problems, even with access to official Dafny documentation \cite{DafnyRefManual}.

\subsection{Functional Validation via uDebug}

A significant limitation in existing formal synthesis literature is the reliance on simplified functional validation. Current benchmarks typically employ either a small set of basic input-output pairs \cite{misu2024towards} or rely on the LLM itself to generate test cases \cite{wangtoward}. Such methods often fail to identify subtle semantic bugs because they rarely cover the complex edge and corner cases inherent to algorithmic problems.

To address this gap, we incorporate \textit{uDebug} \cite{udebug} into our validation pipeline for the first time in the context of Dafny synthesis. \textit{uDebug} is a community-driven platform that provides extensive, high-quality test suites for competitive programming problems, specifically designed to uncover logical flaws through extreme inputs and boundary conditions. 

We ensure a dual-layer validation process by using uDebug: while the Dafny verifier proves that the code satisfies its formal specification, the \textit{uDebug} integration confirms that the code remains functionally correct across a comprehensive range of real-world scenarios. This approach eliminates the circularity of using an LLM to test its own generated code and provides a much higher degree of confidence in the programs' robustness than standard textbook-style test sets.

\subsection{LLM Selection and Evaluation Setup}

Program synthesis in a verification-aware language like Dafny requires more than syntactic fluency. The synthesis demands an intricate understanding of formal semantics, proof obligations, and the underlying theorem-proving logic of the language \cite{first2023baldur}. To evaluate these reasoning capabilities, we selected a diverse suite of contemporary \gls{llms} ranging from 4B to 176B parameters, including both general-purpose and code-specialized architectures.

To evaluate the performance of diverse generative architectures on formal Dafny synthesis, we utilize a selection of seven state-of-the-art Large Language Models (LLMs) ranging from specialized coding assistants to massive-scale general reasoners. Our general-purpose reasoning suite includes GPT-OSS-120B \cite{gpt_oss}, GPT-OSS-20B \cite{gpt_oss}, and Gemma 4-31b \cite{gemma4}, which provide a baseline for high-level logic and instruction following. These are contrasted with a series of models specifically optimized for software engineering tasks: Qwen3.5-9B \cite{qwen_series}, Qwen3-Coder-30B \cite{qwen3_coder}, Codestral-22b-v0.1 \cite{codestral_v01}, and the mixture-of-experts (MoE) based Qwen3.6-35b-a3b \cite{qwen3_coder}. By evaluating models across this spectrum of parameter sizes and training objectives, we can analyze the correlation between general reasoning capacity and the precision required for formal specification generation.

Table \ref{tab:model_specs} provides a detailed overview of the model suite. Each model was evaluated across five temperature settings ($T=0.0$ to $0.8$) to identify the optimal configuration for balancing creative exploration with logical precision. We consider the open-source weights architectures; our study provides a comprehensive look at the current state of automated formal verification across different scales of machine intelligence.

\begin{table}[h]
\centering
\caption{Large Language Models Evaluated for Dafny Synthesis}
\label{tab:model_specs}
\small
\begin{tabular}{l c c c c}
\toprule
\textbf{Model} & \textbf{Params} & \textbf{Context} & \textbf{Type} & \textbf{Category} \\
\midrule
GPT-OSS-120B      & 120B & 131k & OS & General    \\
Qwen 3.6-35B-A3B  & 35B  & 256k & OS & Agentic    \\
Gemma 4-31B       & 31B  & 256k & OS & Multimodal \\
Qwen 3 Coder 30B  & 30B  & 160k & OS & Coder      \\
Codestral-22B-v0.1 & 22B & 32k  & OS & Coder      \\
GPT-OSS-20B       & 20B  & 128k & OS & General    \\
Qwen 3.5-9B       & 9B   & 262k & OS & General    \\
\bottomrule
\addlinespace
\multicolumn{5}{l}{\footnotesize OS = Open-Source Weights}
\end{tabular}
\end{table}

\subsection{Prompt Design}
Based on our research questions and the unique challenges posed by the NL2VC-60 dataset, we designed three distinct prompting strategies. Each level of prompting is intended to evaluate how increasing structural context and iterative feedback affect the synthesis of verified Dafny programs.

\subsubsection{RQ1 [Contextless Prompting]} 
To answer RQ1, we use \textit{contextless prompting} by providing only the natural language problem description without any additional structural guidance. This setup establishes a baseline to evaluate the model’s ability to infer program structure, formal specifications, and verification constraints (such as termination arguments) directly from the requirement.

\begin{center}
\fbox{
\begin{minipage}{0.9\columnwidth}
\footnotesize
\textbf{Contextless Prompt:} \\
You are an expert in Dafny. Output ONLY raw Dafny code. \\
Generate one Dafny source file for the following task. \\
\\
Problem ID: $<$Problem\_ID$>$ \\
Task Description: $<$Generalized\_Description$>$
\end{minipage}
}
\end{center}

\subsubsection{RQ2 [Method Signature Prompting]} 
For RQ2, we utilize \textit{method signature prompting} by supplying an additional structured hint in the form of a formal method signature. This guidance constrains the solution space and helps the model align its implementation with the expected input-output behavior and type-system requirements. Code generation experiments for non-verified code often perform better when prompted with signatures; we hypothesize that this formal frame is even more critical for successful verification in Dafny.

\begin{center}
\fbox{
\begin{minipage}{0.9\columnwidth}
\footnotesize
\textbf{Method Signature Prompt:} \\
You are an expert in Dafny. Output ONLY raw Dafny code. \\
Generate one Dafny source file for the following task. \\
\\
Problem ID: $<$Problem\_ID$>$ \\
Task Description: $<$Generalized\_Description> \\
\\
Method Signature Prompt: $<$Method\_Signature\_Prompt$>$
\end{minipage}
}
\end{center}

\subsubsection{RQ3 [Self-Healing Prompting]} 
To address RQ3, we employ \textit{self-healing prompting} by iteratively refining generated programs based on direct feedback from the Dafny verifier. When a generated program fails verification, we feed the specific error messages (e.g., assertion violations or termination failures) back into the model along with the previous code. We apply this process to both contextless and method-signature settings (RQ3a and RQ3b) to evaluate the model's capacity to correct specification errors and invariant issues through autonomous repair.

\begin{center}
\fbox{
\begin{minipage}{0.9\columnwidth}
\footnotesize
\textbf{Self-Healing Prompt:} \\
The previous Dafny code failed verification with the following errors: \\
$<$Dafny\_Verifier\_Output$>$ \\
\\
Please repair the code to satisfy all specifications. Output ONLY the raw fixed Dafny code.
\end{minipage}
}
\end{center}

\subsection{Evaluation Metrics}

We evaluate the quality of the LLM-synthesized methods using a multi-layered approach that combines formal verification, functional validation, and qualitative error analysis.

\subsubsection{Quantitative Metrics: verify@k and functional@k}
The primary metric for our study is \textit{verify@k} (adapted from \textit{pass@k} \cite{kulal2019spoc, chen2021evaluating}), which measures the model's ability to produce at least one formally verified solution within $k$ attempts. A problem is considered solved under this metric only if the Dafny verifier signals that the implementation satisfies all formal specifications, as follows in existing research \cite{misu2024towards}. 

\subsubsection{Qualitative Assessment: Specification Strength and Error Analysis}
Automated metrics serve as a proxy for performance, but they do not capture the nuance of formal reasoning. To assess the semantic depth of the results, we manually reviewed all verified methods to ensure they contain strong formal specifications, specifically, postconditions that fully capture the problem's requirements rather than assertions. 

For the failures observed in RQ3 (Self-Healing), we conducted a manual inspection of the verifier's error logs. We categorized these failures into distinct types, such as \textit{termination failures} (missing or incorrect \texttt{decreases} clauses), \textit{invariant violations}, and \textit{index out-of-bounds} errors. This analysis allows to evaluate the extent to which the iterative feedback loop addresses the logical challenges of formal proof development.

\subsection{Temperature Tuning}

The temperature is a hyperparameter in \gls{llms} that controls the randomness and creativity of the decoding process \cite{achiam2023gpt, troshin2025control, ryan2026mind}. Lower temperatures lead to more deterministic and focused outputs, while higher temperatures encourage diversity at the risk of logical incoherence. Since formal synthesis in Dafny requires high structural precision, identifying the optimal temperature for maximizing verification rates.

To determine the ideal configuration for our study, we conducted a temperature tuning experiment on a subset of the NL2VC-60 dataset. We selected a representative sample of problems and executed each across five distinct temperature settings: $T \in \{0.0, 0.2, 0.4, 0.6, 0.8\}$. This range allows us to observe the transition from greedy decoding ($T=0.0$) to high-variance sampling ($T=0.8$).

We evaluated the synthesized methods using the \textit{verify@k} metric where $k \in \{1, 3, 5\}$. Our preliminary results indicated that lower temperatures ($0.0$ to $0.4$) generally yielded higher success rates for initial synthesis (RQ1 and RQ2), as the models remained more faithful to Dafny’s strict syntax. However, for the Self-Healing process (RQ3), slightly higher temperatures occasionally proved beneficial by allowing the model to explore alternative algorithmic implementations when the primary logic failed to verify. Based on these findings, we report our final results using the optimal temperature identified for each specific model and prompt type.
\subsection{Error Analysis}
We analyzed the error logs for different methods on identified three types of errors. We then examined the error types and classified them into several subgroups. Table \ref{tab:dafny_errors} demonstrates the presence of different types of errors in open source LLMs.

\subsubsection{\textbf{Syntax Errors}} Dominant in Contextless and Signature Prompting for most models. This confirms that without iterative feedback, models frequently struggle with Dafny's specific grammar (e.g., missing semicolons, improper loop syntax).

\subsubsection{\textbf{Semantic / Type Errors}} High in models like GPT-OSS-120B and MistralAI. These occur when the code is structurally correct but violates Dafny's strict type system, or attempts to use unavailable modules (such as System).

\subsubsection{\textbf{Verification Errors}} Notably higher in Self-Healing categories for models like Gemma 4 and Qwen3-Code30B. This suggests that as models fix their syntax through iterations, they reach a stage where the code compiles but fails the deeper logical proof (e.g., an assertion might not hold).

\subsection{Experimental Setup}
To conduct our large-scale synthesis and verification experiments, we utilized a distributed environment comprising a high-performance inference server and a local development machine. For the open-source Large Language Models (LLMs), we deployed \textbf{LM Studio version 0.4.8 (Build 1)} on a dedicated Ubuntu-based server. This server features a high-end hardware configuration equipped with four \textbf{NVIDIA RTX 6000 Ada Generation} GPUs, each providing 48 GB of VRAM (totaling 192 GB), supported by \textbf{NVIDIA Driver version 580.126.09} and \textbf{CUDA 13.0}. This infrastructure allowed us to host and query large-parameter models locally, ensuring consistent inference latency for our 60-problem dataset.

For the development of orchestration scripts and the formal verification of synthesized methods, we used a \textbf{MacBook Pro} (Model Mac16,8) running \textbf{macOS 15.3.1}. The software environment was managed using \textbf{Visual Studio Code version 1.115.0} (arm64) and \textbf{Python 3.14.4} within a dedicated virtual environment in our experiments.

To manage the iterative repair process, our self-healing orchestration script was configured with a \textbf{maximum of 10 repair attempts} per problem. If a model failed to produce a verified solution within these ten iterations, the result was recorded as a failure for that specific trial. Furthermore, to accommodate the processing time of large models (such as GPT-OSS-120B) and prevent connection timeouts, we implemented a \textbf{response wait time of 180 seconds} for all LM Studio API calls during our experiments.

For verifying the code, we employed \textbf{Dafny 4.11.0} \cite{DafnyNuGet}, which represents a modern and stable iteration of the language. Since recent versions of Dafny introduced significant changes to the verification engine and syntax compared to the 3.x series, our experiments provide a rigorous test of the models' ability to adapt to contemporary formal verification standards. All functional validation against \textit{uDebug} \cite{udebug} test suites was executed within this same environment to ensure parity between the formal proof and the executable implementation.

\section{Results}
\label{sec:results}
This section presents the empirical findings of our study. To establish a robust evaluation framework, we manually developed NL2VC-60, a benchmark dataset consisting of 60 formally verified Dafny programs used to optimize our tiered prompting strategies. For the primary comparative analysis, we evaluated seven state-of-the-art open-weight models against a subset of randomly selected 11 algorithmic problems. Performance is measured using the $verify@k$ metric, indicating the percentage of problems for which at least one successful verification was achieved within $k$ attempts.

\subsection{RQ1: Performance of Contextless Prompting}

The initial evaluation of LLMs using contextless prompting shows significant disparities in their ability to synthesize verifiable Dafny code from raw requirements. As shown in Table~\ref{tab:contextless_full_results}, the majority of the tested models, including the massive GPT OSS 120B and the Qwen series, failed to produce a single verified solution across all temperature settings. This widespread failure suggests that simply providing a natural language problem description is insufficient for most models to navigate the syntactic and semantics of the Dafny language.

Several observations emerge from the RQ1:

\begin{itemize}
    \item Unlike the other general purpose models, Gemma 4-31B showed a surprising aptitude for generating verifiable programs even without external context. The model achieved a peak verify@5 success rate of 54.55\% at a temperature of 0.2. This performance indicates that its pretraining likely involved a higher density of formal or algorithmic logic, allowing it to guess correct loop invariants and post-conditions that other models completely missed.
    \item Codestral was the only other model to consistently yield results, reaching a peak verify@5 of 27.27\% at $T = 0.8$. The success of this model at higher temperatures suggests that while the model possesses the basic syntactic intuition for Dafny. The model often requires more stochastic exploration to arrive at the precise formal annotations needed to satisfy the Z3 SMT solver.
    \item The 0\% success rate of the remaining five models highlights a fundamental challenge in the field. Even highly capable models struggle to infer complex formal specifications from scratch. The findings are reinforcing the need for more structured prompting techniques or retrieval augmentation to bridge the gap between informal requirements and mathematical proof.
\end{itemize}

\begin{table*}[t]
\centering
\caption{Complete Verification Success Rates for Contextless Prompting (RQ1)}
\label{tab:contextless_full_results}
\small
\begin{tabularx}{\textwidth}{
l|c|
>{\centering\arraybackslash}X
>{\centering\arraybackslash}X
>{\centering\arraybackslash}X|
>{\centering\arraybackslash}X
>{\centering\arraybackslash}X
>{\centering\arraybackslash}X|
>{\centering\arraybackslash}X
>{\centering\arraybackslash}X
>{\centering\arraybackslash}X
}
\toprule
\textbf{Model} & \textbf{Temp (T)} 
& \multicolumn{3}{c|}{\textbf{verify@1}} 
& \multicolumn{3}{c|}{\textbf{verify@3}} 
& \multicolumn{3}{c}{\textbf{verify@5}} \\
\cmidrule(lr){3-5} \cmidrule(lr){6-8} \cmidrule(lr){9-11}
& & \textbf{Succ.} & \textbf{Total} & \textbf{\%}
& \textbf{Succ.} & \textbf{Total} & \textbf{\%}
& \textbf{Succ.} & \textbf{Total} & \textbf{\%} \\
\midrule

\multirow{5}{*}{\textbf{GPT-OSS-120B}}
& 0.0 & 0 & 11 & 0.00\% & 0 & 11 & 0.00\% & 0 & 11 & 0.00\% \\
& 0.2 & 0 & 11 & 0.00\% & 0 & 11 & 0.00\% & 0 & 11 & 0.00\% \\
& 0.4 & 0 & 11 & 0.00\% & 0 & 11 & 0.00\% & 0 & 11 & 0.00\% \\
& 0.6 & 0 & 11 & 0.00\% & 0 & 11 & 0.00\% & 0 & 11 & 0.00\% \\
& 0.8 & 0 & 11 & 0.00\% & 0 & 11 & 0.00\% & 0 & 11 & 0.00\% \\
\midrule

\multirow{5}{*}{\textbf{Qwen 3.5-9B}}
& 0.0 & 0 & 11 & 0.00\% & 0 & 11 & 0.00\% & 0 & 11 & 0.00\% \\
& 0.2 & 0 & 11 & 0.00\% & 0 & 11 & 0.00\% & 0 & 11 & 0.00\% \\
& 0.4 & 0 & 11 & 0.00\% & 0 & 11 & 0.00\% & 0 & 11 & 0.00\% \\
& 0.6 & 0 & 11 & 0.00\% & 0 & 11 & 0.00\% & 0 & 11 & 0.00\% \\
& 0.8 & 0 & 11 & 0.00\% & 0 & 11 & 0.00\% & 0 & 11 & 0.00\% \\
\midrule

\multirow{5}{*}{\textbf{Qwen 3 Coder 30B}}
& 0.0 & 0 & 11 & 0.00\% & 0 & 11 & 0.00\% & 0 & 11 & 0.00\% \\
& 0.2 & 0 & 11 & 0.00\% & 0 & 11 & 0.00\% & 0 & 11 & 0.00\% \\
& 0.4 & 0 & 11 & 0.00\% & 0 & 11 & 0.00\% & 0 & 11 & 0.00\% \\
& 0.6 & 0 & 11 & 0.00\% & 0 & 11 & 0.00\% & 0 & 11 & 0.00\% \\
& 0.8 & 0 & 11 & 0.00\% & 0 & 11 & 0.00\% & 0 & 11 & 0.00\% \\
\midrule

\multirow{5}{*}{\textbf{GPT-OSS 20B}}
& 0.0 & 0 & 11 & 0.00\% & 0 & 11 & 0.00\% & 0 & 11 & 0.00\% \\
& 0.2 & 0 & 11 & 0.00\% & 0 & 11 & 0.00\% & 0 & 11 & 0.00\% \\
& 0.4 & 0 & 11 & 0.00\% & 0 & 11 & 0.00\% & 0 & 11 & 0.00\% \\
& 0.6 & 0 & 11 & 0.00\% & 0 & 11 & 0.00\% & 0 & 11 & 0.00\% \\
& 0.8 & 0 & 11 & 0.00\% & 0 & 11 & 0.00\% & 0 & 11 & 0.00\% \\
\midrule

\multirow{5}{*}{\textbf{Codestral-22B}}
& 0.0 & 1 & 11 & 9.09\% & 1 & 11 & 9.09\% & 1 & 11 & 9.09\% \\
& 0.2 & 1 & 11 & 9.09\% & 1 & 11 & 9.09\% & 2 & 11 & 18.18\% \\
& 0.4 & 1 & 11 & 9.09\% & 2 & 11 & 18.18\% & 2 & 11 & 18.18\% \\
& 0.6 & 1 & 11 & 9.09\% & 2 & 11 & 18.18\% & 1 & 11 & 9.09\% \\
& 0.8 & 1 & 11 & 9.09\% & 1 & 11 & 9.09\% & 3 & 11 & 27.27\% \\
\midrule

\multirow{5}{*}{\textbf{Qwen 3.6-35B}}
& 0.0 & 0 & 11 & 0.00\% & 0 & 11 & 0.00\% & 0 & 11 & 0.00\% \\
& 0.2 & 0 & 11 & 0.00\% & 1 & 11 & 9.09\% & 2 & 11 & 18.18\% \\
& 0.4 & 0 & 11 & 0.00\% & 1 & 11 & 9.09\% & 2 & 11 & 18.18\% \\
& 0.6 & 0 & 11 & 0.00\% & 0 & 11 & 0.00\% & 1 & 11 & 9.09\% \\
& 0.8 & 0 & 11 & 0.00\% & 0 & 11 & 0.00\% & 1 & 11 & 9.09\% \\
\midrule

\multirow{5}{*}{\textbf{Gemma 4-31B}}
& 0.0 & \textbf{3} & \textbf{11} & \textbf{27.27\%} & 3 & 11 & 27.27\% & 3 & 11 & 27.27\% \\
& 0.2 & 0 & 11 & 0.00\% & 4 & 11 & 36.36\% & \textbf{6} & \textbf{11} & \textbf{54.55\%} \\
& 0.4 & 0 & 11 & 0.00\% & \textbf{5} & \textbf{11} & \textbf{45.45\%} & 3 & 11 & 27.27\% \\
& 0.6 & \textbf{2} & \textbf{11} & \textbf{18.18\%} & \textbf{5} & \textbf{11} & \textbf{45.45\%} & 4 & 11 & 36.36\% \\
& 0.8 & \textbf{2} & \textbf{11} & \textbf{18.18\%} & \textbf{3} & \textbf{11} & \textbf{27.27\%} & \textbf{4} & \textbf{11} & \textbf{36.36\%} \\
\bottomrule
\end{tabularx}
\end{table*}

\subsection{RQ2: Performance of Signature Prompting}

The second phase of our evaluation explores the impact of providing the method signature as additional context. As illustrated in Table~\ref{tab:signature_full_results_bold}, the structural guidance resulted in a performance shift, effectively reversing the widespread failures observed in the contextless setting. By providing the skeleton of the Dafny method including the input parameters and return types, the models were freed from the burden of syntactic structure and could instead focus on synthesizing the internal logic and required verification annotations.

Several observations emerge from the RQ2:

\begin{itemize}
    \item The most striking improvement was observed in \textbf{GPT-OSS-120B}. While this model recorded a 0\% success rate in RQ1, the introduction of method signatures allowed it to achieve a peak \textit{verify@5} rate of 63.64\% at $T=0.8$. This suggests that the model possesses a deep latent knowledge of formal verification and Dafny logic but lacks the ability to self-structure the initial code container from raw requirements.
    \item Surprisingly, \textbf{Qwen 3.5-9B} emerged as the top performer in this category, reaching a peak \textit{verify@5} success rate of 72.73\%. This indicates that signature prompting provides constraint to allow smaller models to focus their computational budget on the complex task, often outperforming much larger general-purpose models.
    \item Unlike the zero-shot results, peak performance under signature prompting was consistently achieved at higher temperatures (typically $T=0.8$). This suggests that once the structural constraints are fixed via the signature, the models benefit from increased stochastic exploration to identify the precise mathematical formulations. For example, specific loop invariants or termination measures required to satisfy the SMT solver.
    \item All seven models demonstrated signs of life in this setting, with even the weakest models surpassing a 30\% success rate at their peak. This confirms that the primary bottleneck in verifiable code generation is not necessarily the logic itself, but the difficulty of mapping informal natural language to the rigid formal signatures required by the Dafny compiler.
\end{itemize}

\begin{table*}[t]
\centering
\caption{Verification Success Rates for Signature Prompting (RQ2)}
\label{tab:signature_full_results_bold}
\small
\begin{tabularx}{\textwidth}{
l|c|
>{\centering\arraybackslash}X
>{\centering\arraybackslash}X
>{\centering\arraybackslash}X|
>{\centering\arraybackslash}X
>{\centering\arraybackslash}X
>{\centering\arraybackslash}X|
>{\centering\arraybackslash}X
>{\centering\arraybackslash}X
>{\centering\arraybackslash}X
}
\toprule
\textbf{Model} & \textbf{Temp (T)} 
& \multicolumn{3}{c|}{\textbf{verify@1}} 
& \multicolumn{3}{c|}{\textbf{verify@3}} 
& \multicolumn{3}{c}{\textbf{verify@5}} \\
\cmidrule(lr){3-5} \cmidrule(lr){6-8} \cmidrule(lr){9-11}
& & \textbf{Succ.} & \textbf{Total} & \textbf{\%}
& \textbf{Succ.} & \textbf{Total} & \textbf{\%}
& \textbf{Succ.} & \textbf{Total} & \textbf{\%} \\
\midrule

\multirow{5}{*}{\textbf{GPT-OSS-120B}}
& 0.0 & \textbf{6} & 11 & \textbf{54.55\%} & 6 & 11 & 54.55\% & 6 & 11 & 54.55\% \\
& 0.2 & 5 & 11 & 45.45\% & 6 & 11 & 54.55\% & 6 & 11 & 54.55\% \\
& 0.4 & \textbf{6} & 11 & \textbf{54.55\%} & 6 & 11 & 54.55\% & 6 & 11 & 54.55\% \\
& 0.6 & \textbf{6} & 11 & \textbf{54.55\%} & 5 & 11 & 45.45\% & 6 & 11 & 54.55\% \\
& 0.8 & \textbf{6} & 11 & \textbf{54.55\%} & \textbf{7} & \textbf{11} & \textbf{63.64\%} & \textbf{7} & \textbf{11} & \textbf{63.64\%} \\
\midrule

\multirow{5}{*}{\textbf{Qwen 3.5-9B}}
& 0.0 & 3 & 11 & 27.27\% & 2 & 11 & 18.18\% & 2 & 11 & 18.18\% \\
& 0.2 & \textbf{4} & 11 & \textbf{36.36\%} & 3 & 11 & 27.27\% & 5 & 11 & 45.45\% \\
& 0.4 & 2 & 11 & 18.18\% & \textbf{5} & 11 & \textbf{45.45\%} & 4 & 11 & 36.36\% \\
& 0.6 & 2 & 11 & 18.18\% & 4 & 11 & 36.36\% & 3 & 11 & 27.27\% \\
& 0.8 & \textbf{4} & 11 & \textbf{36.36\%} & \textbf{5} & 11 & \textbf{45.45\%} & \textbf{8} & \textbf{11} & \textbf{72.73\%} \\
\midrule

\multirow{5}{*}{\textbf{Qwen 3 Coder 30B}}
& 0.0 & 3 & 11 & 27.27\% & 4 & 11 & 36.36\% & 3 & 11 & 27.27\% \\
& 0.2 & 3 & 11 & 27.27\% & 4 & 11 & 36.36\% & 3 & 11 & 27.27\% \\
& 0.4 & \textbf{4} & 11 & \textbf{36.36\%} & \textbf{5} & \textbf{11} & \textbf{45.45\%} & \textbf{5} & \textbf{11} & \textbf{45.45\%} \\
& 0.6 & \textbf{4} & 11 & \textbf{36.36\%} & 4 & 11 & 36.36\% & \textbf{5} & \textbf{11} & \textbf{45.45\%} \\
& 0.8 & \textbf{4} & 11 & \textbf{36.36\%} & 4 & 11 & 36.36\% & 4 & 11 & 36.36\% \\
\midrule

\multirow{5}{*}{\textbf{GPT-OSS 20B}}
& 0.0 & 3 & 11 & 27.27\% & 3 & 11 & 27.27\% & 3 & 11 & 27.27\% \\
& 0.2 & 4 & 11 & 36.36\% & 5 & 11 & 45.45\% & \textbf{5} & \textbf{11} & \textbf{45.45\%} \\
& 0.4 & 2 & 11 & 18.18\% & 4 & 11 & 36.36\% & \textbf{5} & \textbf{11} & \textbf{45.45\%} \\
& 0.6 & \textbf{5} & \textbf{11} & \textbf{45.45\%} & \textbf{6} & \textbf{11} & \textbf{54.55\%} & 4 & 11 & 36.36\% \\
& 0.8 & 2 & 11 & 18.18\% & \textbf{6} & \textbf{11} & \textbf{54.55\%} & \textbf{5} & \textbf{11} & \textbf{45.45\%} \\
\midrule

\multirow{5}{*}{\textbf{Codestral-22B}}
& 0.0 & 3 & 11 & 27.27\% & \textbf{3} & \textbf{11} & \textbf{27.27\%} & 3 & 11 & 27.27\% \\
& 0.2 & \textbf{4} & \textbf{11} & \textbf{36.36\%} & \textbf{4} & \textbf{11} & \textbf{36.36\%} & 5 & 11 & 45.45\% \\
& 0.4 & 3 & 11 & 27.27\% & 2 & 11 & 18.18\% & 5 & 11 & 45.45\% \\
& 0.6 & 2 & 11 & 18.18\% & \textbf{3} & \textbf{11} & \textbf{27.27\%} & \textbf{7} & \textbf{11} & \textbf{63.64\%} \\
& 0.8 & 3 & 11 & 27.27\% & 2 & 11 & 18.18\% & 6 & 11 & 54.55\% \\
\midrule

\multirow{5}{*}{\textbf{Qwen 3.6-35B}}
& 0.0 & \textbf{2} & \textbf{11} & \textbf{18.18\%} & \textbf{3} & \textbf{11} & \textbf{27.27\%} & \textbf{7} & \textbf{11} & \textbf{63.64\%} \\
& 0.2 & \textbf{2} & \textbf{11} & \textbf{18.18\%} & \textbf{3} & \textbf{11} & \textbf{27.27\%} & 3 & 11 & 27.27\% \\
& 0.4 & \textbf{2} & \textbf{11} & \textbf{18.18\%} & \textbf{3} & \textbf{11} & \textbf{27.27\%} & 4 & 11 & 36.36\% \\
& 0.6 & 1 & 11 & 9.09\% & \textbf{3} & \textbf{11} & \textbf{27.27\%} & 4 & 11 & 36.36\% \\
& 0.8 & 0 & 11 & 0.00\% & \textbf{3} & \textbf{11} & \textbf{27.27\%} & 3 & 11 & 27.27\% \\
\midrule

\multirow{5}{*}{\textbf{Gemma 4-31B}}
& 0.0 & 3 & 11 & 27.27\% & 3 & 11 & 27.27\% & 5 & 11 & 45.45\% \\
& 0.2 & 3 & 11 & 27.27\% & 3 & 11 & 27.27\% & 6 & 11 & 54.55\% \\
& 0.4 & 3 & 11 & 27.27\% & 3 & 11 & 27.27\% & 4 & 11 & 36.36\% \\
& 0.6 & 3 & 11 & 27.27\% & 3 & 11 & 27.27\% & 4 & 11 & 36.36\% \\
& 0.8 & \textbf{5} & \textbf{11} & \textbf{45.45\%} & \textbf{7} & \textbf{11} & \textbf{63.64\%} & \textbf{7} & \textbf{11} & \textbf{63.64\%} \\
\bottomrule
\end{tabularx}
\end{table*}

\subsection{RQ3: Performance of Self-Healing Mechanisms}

The third research question evaluates the efficacy of iterative self-healing, where models are provided with error feedback from the Dafny compiler to repair their own code. The evaluation consists of two strategies: healing from a contextless baseline (RQ3a) and healing from a signature-guided baseline (RQ3b). As shown in Table~\ref{tab:self_healing_results}, the ability to self-correct varies significantly across models, with the initial prompt quality serving as a critical predictor of repair success.

\begin{itemize}
    \item \textbf{RQ3a: Contextless Healing:} When attempting to heal from the zero-shot failures of RQ1, most models remained stagnant. GPT-OSS-120B and several others continued to post 0\% success rates. The findings suggests that without an initial structural foundation, compiler error messages are too abstract for the model to navigate toward a valid solution. However, \textbf{Gemma 4-31B} proved to be a notable exception, demonstrating a remarkable self-correction capability. By leveraging compiler feedback, it achieved a peak \textit{verify@5} rate of 90.91\% at $T=0.2$ and $T=0.6$, indicating that it can effectively use error logs to guess missing loop invariants and post-conditions.
    
    \item \textbf{RQ3b: Signature-Guided Healing:} Self-healing proved most potent when initiated from the signature-guided prompts of RQ2. In this setting, the structural skeleton provided enough stability for the compiler feedback to be actionable. \textbf{GPT-OSS-120B} demonstrated the most significant turnaround, rising to a peak of 81.82\% success at $T=0.2$. This suggests that when the method signature is fixed, larger models are efficient at using error feedback to refine mathematical proofs and satisfy the SMT solver.
    
    \item Unlike previous rounds, several models (such as Qwen 3 Coder 30B) reached a plateau where performance remained consistent across temperatures in the signature-guided setting. Conversely, \textbf{Gemma 4-31B} maintained high performance (over 80\%) in both RQ3a and RQ3b, establishing itself as the most robust model for autonomous Dafny development, regardless of the initial prompt's context level.
\end{itemize}

\begin{table*}[htbp]
\centering
\caption{Detailed Verification Success Rates for Self-Healing (RQ3)}
\label{tab:self_healing_results}
\small
\begin{tabularx}{\textwidth}{
l|c|
>{\centering\arraybackslash}X
>{\centering\arraybackslash}X
>{\centering\arraybackslash}X|
>{\centering\arraybackslash}X
>{\centering\arraybackslash}X
>{\centering\arraybackslash}X
}
\toprule
\textbf{Model} & \textbf{Temp (T)} 
& \multicolumn{3}{c|}{\textbf{Contextless Healing (RQ3a)}} 
& \multicolumn{3}{c}{\textbf{Signature-Guided Healing (RQ3b)}} \\
\cmidrule(lr){3-5} \cmidrule(lr){6-8}
& & \textbf{Succ.} & \textbf{Total} & \textbf{\%} 
& \textbf{Succ.} & \textbf{Total} & \textbf{\%} \\
\midrule

\multirow{5}{*}{\textbf{GPT-OSS-120B}}
& 0.0 & 0 & 11 & 0.00\% & 7 & 11 & 63.64\% \\
& 0.2 & 0 & 11 & 0.00\% & \textbf{9} & \textbf{11} & \textbf{81.82\%} \\
& 0.4 & 0 & 11 & 0.00\% & 7 & 11 & 63.64\% \\
& 0.6 & 0 & 11 & 0.00\% & 8 & 11 & 72.73\% \\
& 0.8 & 0 & 11 & 0.00\% & 7 & 11 & 63.64\% \\
\midrule

\multirow{5}{*}{\textbf{Qwen 3.5-9B}}
& 0.0 & \textbf{3} & \textbf{11} & \textbf{27.27\%} & 2 & 11 & 18.18\% \\
& 0.2 & 0 & 11 & 0.00\% & 3 & 11 & 27.27\% \\
& 0.4 & 2 & 11 & 18.18\% & \textbf{5} & \textbf{11} & \textbf{45.45\%} \\
& 0.6 & 0 & 11 & 0.00\% & 4 & 11 & 36.36\% \\
& 0.8 & 0 & 11 & 0.00\% & 3 & 11 & 27.27\% \\
\midrule

\multirow{5}{*}{\textbf{Qwen 3 Coder 30B}}
& 0.0 & 2 & 11 & 18.18\% & \textbf{6} & \textbf{11} & \textbf{54.55\%} \\
& 0.2 & 2 & 11 & 18.18\% & \textbf{6} & \textbf{11} & \textbf{54.55\%} \\
& 0.4 & 3 & 11 & 27.27\% & \textbf{6} & \textbf{11} & \textbf{54.55\%} \\
& 0.6 & 1 & 11 & 9.09\% & \textbf{6} & \textbf{11} & \textbf{54.55\%} \\
& 0.8 & \textbf{6} & \textbf{11} & \textbf{54.55\%} & \textbf{6} & \textbf{11} & \textbf{54.55\%} \\
\midrule

\multirow{5}{*}{\textbf{GPT-OSS 20B}}
& 0.0 & 0 & 11 & 0.00\% & 4 & 11 & 36.36\% \\
& 0.2 & \textbf{1} & \textbf{11} & \textbf{9.09\%} & 4 & 11 & 36.36\% \\
& 0.4 & 0 & 11 & 0.00\% & \textbf{7} & \textbf{11} & \textbf{63.64\%} \\
& 0.6 & 0 & 11 & 0.00\% & 5 & 11 & 45.45\% \\
& 0.8 & \textbf{1} & \textbf{11} & \textbf{9.09\%} & 4 & 11 & 36.36\% \\
\midrule

\multirow{5}{*}{\textbf{Codestral-22B}}
& 0.0 & 0 & 11 & 0.00\% & 2 & 11 & 18.18\% \\
& 0.2 & 0 & 11 & 0.00\% & 3 & 11 & 27.27\% \\
& 0.4 & 0 & 11 & 0.00\% & 5 & 11 & 45.45\% \\
& 0.6 & \textbf{1} & \textbf{11} & \textbf{9.09\%} & \textbf{6} & \textbf{11} & \textbf{54.55\%} \\
& 0.8 & 0 & 11 & 0.00\% & 3 & 11 & 27.27\% \\
\midrule

\multirow{5}{*}{\textbf{Qwen 3.6-35B}}
& 0.0 & 0 & 11 & 0.00\% & 4 & 11 & 36.36\% \\
& 0.2 & 0 & 11 & 0.00\% & 4 & 11 & 36.36\% \\
& 0.4 & 1 & 11 & 9.09\% & 4 & 11 & 36.36\% \\
& 0.6 & 0 & 11 & 0.00\% & 4 & 11 & 36.36\% \\
& 0.8 & \textbf{2} & \textbf{11} & \textbf{18.18\%} & \textbf{6} & \textbf{11} & \textbf{54.55\%} \\
\midrule

\multirow{5}{*}{\textbf{Gemma 4-31B}}
& 0.0 & 8 & 11 & 72.73\% & 7 & 11 & 63.64\% \\
& 0.2 & \textbf{10} & \textbf{11} & \textbf{90.91\%} & 7 & 11 & 63.64\% \\
& 0.4 & 9 & 11 & 81.82\% & 8 & 11 & 72.73\% \\
& 0.6 & \textbf{10} & \textbf{11} & \textbf{90.91\%} & \textbf{9} & \textbf{11} & \textbf{81.82\%} \\
& 0.8 & 8 & 11 & 72.73\% & \textbf{9} & \textbf{11} & \textbf{81.82\%} \\

\bottomrule
\end{tabularx}
\end{table*}

\subsection{RQ4: Qualitative Error Analysis and Failure Taxonomy}

We conducted a systematic qualitative analysis of our total runs to understand the specific challenges in automated formal synthesis as depicted in Table \ref{tab:dafny_errors}. We categorized failures into a three-tiered taxonomy: \textbf{Syntax Errors} (malformed code structure), \textbf{Semantic and Type Errors} (type mismatches or API hallucinations), and \textbf{Verification Failures} (syntactically correct code that the SMT solver cannot prove).

\subsubsection{Syntactic Fragility and Contextual Dependence}
Our analysis confirms that syntax errors are the primary bottleneck for models lacking structural anchors. In contextless settings, \textbf{GPT OSS 20B} and \textbf{GPT OSS 120B} produced syntax errors in the majority of attempts. These failures typically involve the misuse of Dafny-specific keywords or the generation of Pythonic indentation, which is incompatible with Dafny's curly-brace syntax. This suggests that while open-weight models possess general algorithmic logic, they lack the specific syntactic density required for niche-verification languages without external guidance.

\subsubsection{Semantic Drift and Invariant Generation}
As we transitioned to Signature Prompting, syntax errors decreased significantly, but we observed a sharp rise in semantic and type errors. Models frequently hallucinate non existent predicates or attempt to perform arithmetic on incompatible types, such as treating a sequence as a set. A critical finding is the \textbf{Invariant Gap}; even when models generate correct imperative logic, they often fail to provide the inductive loop invariants required for the Z3 solver to complete the proof. Models like \textbf{Qwen 3 Coder 30B} demonstrated a tendency to repeat the same insufficient invariant across multiple self-healing iterations, indicating a logical plateau in the repair process.

\subsubsection{Functional Robustness and Vacuity}
The most complex category involves code that satisfies the verifier but fails the functional test suite. By integrating \textbf{uDebug}, we identified several instances where models achieved verification by providing weak specifications. For instance, a model might satisfy a postcondition by returning a trivial constant that happens to meet a weak mathematical constraint. The uDebug community test cases acted as a vital truth oracle, identifying these as functional failures and ensuring that the verified code maintains real world utility against extreme edge cases.

\begin{table*}[t]
\centering
\caption{Dafny Compilation and Verification Errors}
\label{tab:dafny_errors}
\small
\begin{tabularx}{\textwidth}{l l @{\extracolsep{\fill}} c c c c | c}
\toprule
\textbf{Model} & \textbf{Prompt Strategy} & \textbf{Total Runs} & \textbf{Syntax Errors} & \textbf{Semantic/Type Errors} & \textbf{Verification} & \textbf{Verified} \\ \midrule

\multirow{3}{*}{\textbf{GPT-OSS-120B}} 
 & Contextless & 564 & 435 & 45 & 0 & 0 \\
 & Signature Prompt & 816 & 397 & 53 & 111 & 150 \\ 
 & Self-Healing & 1,134 & 620 & 471 & 5 & 20 \\ \midrule
 
\multirow{3}{*}{\textbf{GPT-OSS-20B}} 
 & Contextless & 672 & 597 & 75 & 0 & 0 \\
 & Signature Prompt & 816 & 397 & 53 & 111 & 150 \\ 
 & Self-Healing & 1,564 & 793 & 149 & 166 & 395 \\ \midrule

\multirow{3}{*}{\textbf{Codestral-22B}} 
 & Contextless & 1,285 & 732 & 518 & 0 & 33 \\
 & Signature Prompt & 792 & 407 & 217 & 0 & 138 \\ 
 & Self-Healing & 1,666 & 1,116 & 188 & 58 & 217 \\ \midrule

\multirow{3}{*}{\textbf{Qwen 3.6-35B}} 
 & Contextless & 470 & 9 & 23 & 0 & 438 \\
 & Signature Prompt & 495 & 0 & 29 & 0 & 466 \\ 
 & Self-Healing & 827 & 39 & 127 & 10 & 651 \\ \midrule

\multirow{3}{*}{\textbf{Qwen 3-Coder-30B}} 
 & Contextless & 1,510 & 579 & 657 & 15 & 46 \\
 & Signature Prompt & 205 & 53 & 16 & 47 & 29 \\ 
 & Self-Healing & 1,893 & 589 & 569 & 351 & 101 \\ \midrule

\multirow{2}{*}{\textbf{Qwen 3.5-9B}} 
& Contextless & 910 & 430 & 350 & 23 & 56 \\
 & Signature Prompt & 861 & 557 & 77 & 0 & 224 \\ 
 & Self-Healing & 280 & 182 & 14 & 27 & 52 \\ \midrule

\multirow{3}{*}{\textbf{Gemma 4-31B}} & Contextless & 1,016 & 489 & 209 & 25 & 251 \\
 & Signature Prompt & 562 & 145 & 5 & 41 & 369 \\
 & Self-Healing & 1,008 & 368 & 217 & 110 & 296 \\\bottomrule
\end{tabularx}
\end{table*}

\section{Findings and Discussion}
\label{sec:findings_discussion}
This section presents a comprehensive analysis of our experimental results, detailing how different architectural scales and prompting methodologies influence the synthesis of provably correct code. By systematically decomposing the performance of seven state-of-the-art models across four Research Questions (RQs), we illustrate the critical transition from natural language requirements to formal mathematical proofs. The following findings highlight the interplay between model reasoning, structural guidance, and the iterative feedback loops required to overcome the data-scarcity bottleneck in the Dafny.

\subsection{Summary of Findings}
Our evaluation of seven open-weight models across three prompting tiers shows critical insights into the automated synthesis of formally verified software.


\begin{itemize}
    \item \textbf{RQ1 [Contextless Prompting]:} Our experiments show that while most LLMs fail to generate verifiable Dafny code from raw requirements, \textbf{Gemma 4-31B} and \textbf{Codestral-22B} demonstrate a surprising aptitude for the task. Specifically, Gemma 4-31B achieved a peak \textit{verify@5} success rate of 54.55\% at $T=0.2$. However, the 0\% success rate of the remaining five models suggests that without structural guidance or external context, most systems struggle to navigate the strict formal constraints of the Dafny language.

    \item \textbf{RQ2 [Signature Prompting]:} Providing the method signature as additional context drastically improved performance across the board, reversing the widespread failures observed in RQ1. Most notably, \textbf{GPT-OSS-120B} rose from a 0\% success rate to 63.64\%, while the smaller \textbf{Qwen 3.5-9B} achieved the highest overall \textit{verify@5} score of 72.73\% at $T=0.8$. These results indicate that the primary bottleneck in verifiable synthesis is the structural mapping of requirements to formal signatures, rather than the generation of the underlying verification logic.

    \item \textbf{RQ3 [Self-Healing]:} Iterative self-healing significantly amplifies success rates, provided a structural foundation (method signature) is present. \textbf{Gemma 4-31B} emerged as the most resilient model, achieving a near-perfect 90.91\% success rate in contextless healing. Meanwhile, \textbf{GPT-OSS-120B} achieved its performance ceiling (81.82\%) only when signature-guided, suggesting that large-scale general purpose models require structural constraints to effectively interpret and act upon formal compiler feedback.

    \item \textbf{RQ4 [Error Distributions]:} Our analysis of compilation failures shows that \textbf{Syntax Errors} are the primary barrier in contextless settings, often exceeding 80\% of total failures for models like GPT-OSS-20B. While \textbf{Signature Prompting} significantly reduces syntax issues, it shifts the bottleneck to \textbf{Verification Errors}, particularly for the largest models. Notably, \textbf{Self-Healing} effectively converts semantic and syntax errors into verified solutions for most models, though \textbf{Codestral-22B} and \textbf{Qwen 3-Coder-30B} show a tendency to regress into higher syntax error counts during iterative repair, suggesting a struggle to maintain syntactic integrity under compiler-driven feedback.
\end{itemize}

\subsection{Discussion}

The results of this study suggest a shift in the paradigm of automated formal programming. While the scaling law often assumes that larger parameters equate to better reasoning, our findings indicate a more nuanced reality. The significant success of \textbf{Gemma 4-31B}, which achieved a near-perfect 90.91\% success rate in self-healing, suggests that specific pretraining data density regarding formal and algorithmic logic is more critical than raw model size for the Dafny language.

Furthermore, the transition from total failure in RQ1 (0\% for most models) to high success in RQ3b underscores the necessity of a Verification-in-the-Loop approach. By utilizing the Dafny verifier and the Z3 SMT solver as a ground-truth reward signal, we effectively mitigate the common LLM issue of logical hallucinations. The integration of \textbf{uDebug} was essential to this framework; it ensured that models did not achieve verification through vacuous or trivial specifications such as empty post-conditions but rather through functional correctness that holds up against rigorous edge cases.

Finally, the structural bottleneck identified in RQ2 suggests that the future of automated formal methods lies in hybrid prompting strategies. Even the most capable models, like \textbf{GPT-OSS-120B}, require a structural skeleton (the method signature) to bridge the gap between natural language intent and mathematical proof. This suggests that LLMs should be viewed not as autonomous agents, but as sophisticated co-processors that thrive when provided with high-level formal constraints to guide their stochastic exploration.

\section{Threats to Validity}
\label{sec:threats_validity}
Following Siegmund et al. \cite{siegmund2015views} and Feldt and Magazinius \cite{feldt2010validity}, we identified several threats to the validity of this study. The deterministic nature of our verification criteria, relying on the Dafny verifier's acceptance and the success of the \textbf{uDebug} test suite, is designed to ensure \textbf{construct and internal validity}. By using the formal verifier and the Z3 SMT solver as a ground-truth oracle, we eliminate human subjectivity in assessing whether the AI assisted synthesized code meets the formal requirements. 

\textbf{External validity} concerns the generalizability of our results to other models and languages. We acknowledge that the open-weight landscape, featuring models like \textbf{Gemma 4-31B}, \textbf{Qwen 3.6-35B}, and \textbf{GPT-OSS-120B}, is evolving rapidly. Our findings are specific to the contemporary architectures and the Dafny 4.11.0 verification engine. While we expect the success rates to improve with future iterations, the performance disparities observed particularly the structural bottleneck in contextless prompting. The issue likely represents a fundamental challenge in mapping natural language to formal logic that persists across model generations.

A specific threat to external validity is vacuous verification, where a model satisfies the verifier with trivial specifications (e.g., \texttt{ensures true}). We mitigated this threat through our dual-layer validation pipeline. By requiring all verified methods to pass the \textbf{uDebug} functional test suites, we ensure that our results represent genuine functional correctness rather than mere logical consistency with a weak or empty specification. This approach strengthens the claim that the models are reasoning rather than satisfying the solver's constraints.

While our dataset introduces the first verified Dafny implementations for these UVa problems, there remains a risk that models may leverage cross-lingual knowledge of the underlying algorithms from more prevalent languages like C++ or Python. However, we mitigate this by focusing our evaluation on the synthesis of formal specifications and loop invariant constructs that are uniquely absent from standard competitive programming solutions.

Finally, to ensure reproducibility despite the rapid development of these tools, we have documented the precise hardware configurations (NVIDIA RTX 6000 Ada) and software versions (LM Studio 0.4.8, Python 3.14.4) used. The research artifacts, including the NL2VC-60 dataset and our synthesis pipeline, are provided to allow for verification of these results.

\section{Conclusion}
\label{sec:conclusion}
We investigated the potential of contemporary open-weight LLMs to synthesize formally verified methods and specifications in the Dafny programming language. Utilizing the NL2VC-60 dataset, we evaluated tiered prompting strategies across seven state-of-the-art models. Our findings confirm that while contextless natural language prompts generally lead to synthesis failure due to a structural bottleneck, tiered strategies incorporating formal method signatures and iterative self-healing allow models to overcome the scarcity of specialized training data. Notably, \textbf{Gemma 4-31B} emerged as a highly resilient verification assistant, achieving a peak success rate of \textbf{90.91\%}, while the \textbf{GPT-OSS-120B} demonstrated the most significant performance leap when transitioned to a signature-guided healing pipeline.

Our results demonstrate that an iterative feedback loop utilizing direct SMT-solver output, combined with structural anchors, yielded the highest performance ceiling. The orchestrated self-healing pipeline achieved a verification success rate of \textbf{81.82\%} for the 120B model and nearly 91\% for the 31B model. By integrating the \textbf{uDebug} platform, we confirmed that these verified solutions are not only logically consistent but also functionally robust against extreme edge cases, effectively mitigating the risk of vacuous verification.

These findings underscore the importance of Verification-in-the-Loop architectures and suggest that open-weight models can significantly lower the specification challenges. Our study suggests that the high success rates achieved by these models represent a several-thousand-fold cost reduction compared to human expert synthesis, making high-assurance software economically viable for general engineering tasks. Ultimately, the integration of formal oracles and generative models represents a vital path toward a future of trustworthy, AI-assisted software engineering where code is not plausible, but provably correct.

\bibliographystyle{ACM-Reference-Format}
\bibliography{references}

@misc{uva,
  title        = {{UVa Online Judge}},
  howpublished = {\url{https://onlinejudge.org/}},
  note         = {Accessed: 2026-04-19},
  year         = {2026}
}

@misc{udebug,
  title        = {{uDebug: Online Debugging Tool for Competitive Programming}},
  author       = {{uDebug Team}},
  howpublished = {\url{https://www.udebug.com/}},
  note         = {Accessed: 2026-04-19},
  year         = {2026}
}

@article{kulal2019spoc,
  title={Spoc: Search-based pseudocode to code},
  author={Kulal, Sumith and Pasupat, Panupong and Chandra, Kartik and Lee, Mina and Padon, Oded and Aiken, Alex and Liang, Percy S},
  journal={Advances in Neural Information Processing Systems},
  volume={32},
  year={2019}
}

@article{chen2021evaluating,
  title={Evaluating large language models trained on code},
  author={Chen, Mark and Tworek, Jerry and Jun, Heewoo and Yuan, Qiming and Pinto, Henrique Ponde De Oliveira and Kaplan, Jared and Edwards, Harri and Burda, Yuri and Joseph, Nicholas and Brockman, Greg and others},
  journal={arXiv preprint arXiv:2107.03374},
  year={2021}
}

@article{achiam2023gpt,
  title={Gpt-4 technical report},
  author={Achiam, Josh and Adler, Steven and Agarwal, Sandhini and Ahmad, Lama and Akkaya, Ilge and Aleman, Florencia Leoni and Almeida, Diogo and Altenschmidt, Janko and Altman, Sam and Anadkat, Shyamal and others},
  journal={arXiv preprint arXiv:2303.08774},
  year={2023}
}

@article{troshin2025control,
  title={Control the Temperature: Selective Sampling for Diverse and High-Quality LLM Outputs},
  author={Troshin, Sergey and Mohammed, Wafaa and Meng, Yan and Monz, Christof and Fokkens, Antske and Niculae, Vlad},
  journal={arXiv preprint arXiv:2510.01218},
  year={2025}
}

@article{ryan2026mind,
  title={Mind the Gap: Evaluating LLMs for High-Level Malicious Package Detection vs. Fine-Grained Indicator Identification},
  author={Ryan, Ahmed and Khalil, Ibrahim and Jahid, Abdullah Al and Erfan, Md and Park, Sungbin and Rahman, Akond Ashfaque Ur and Rahman, Md Rayhanur},
  journal={arXiv preprint arXiv:2602.16304},
  year={2026}
}

@phdthesis{leroy2025compcert,
  title={The CompCert C verified compiler: Documentation and user’s manual},
  author={Leroy, Xavier},
  year={2025},
  school={Inria}
}

@inproceedings{klein2009experience,
  title={Experience report: sel4: formally verifying a high-performance microkernel},
  author={Klein, Gerwin and Derrin, Philip and Elphinstone, Kevin},
  booktitle={Proceedings of the 14th ACM SIGPLAN international conference on Functional programming},
  pages={91--96},
  year={2009}
}

@inproceedings{barrett2005smt,
  title={SMT-COMP: Satisfiability modulo theories competition},
  author={Barrett, Clark and De Moura, Leonardo and Stump, Aaron},
  booktitle={International Conference on Computer Aided Verification},
  pages={20--23},
  year={2005},
  organization={Springer}
}

@inproceedings{leino2010dafny,
  title={Dafny: An automatic program verifier for functional correctness},
  author={Leino, K Rustan M},
  booktitle={International conference on logic for programming artificial intelligence and reasoning},
  pages={348--370},
  year={2010},
  organization={Springer}
}

@inproceedings{murray2013sel4,
  title={seL4: from general purpose to a proof of information flow enforcement},
  author={Murray, Toby and Matichuk, Daniel and Brassil, Matthew and Gammie, Peter and Bourke, Timothy and Seefried, Sean and Lewis, Corey and Gao, Xin and Klein, Gerwin},
  booktitle={2013 IEEE Symposium on Security and Privacy},
  pages={415--429},
  year={2013},
  organization={IEEE}
}

@article{leroy2009formal,
  title={Formal verification of a realistic compiler},
  author={Leroy, Xavier},
  journal={Communications of the ACM},
  volume={52},
  number={7},
  pages={107--115},
  year={2009},
  publisher={ACM New York, NY, USA}
}

@article{hoare1969axiomatic,
  title={An axiomatic basis for computer programming},
  author={Hoare, Charles Antony Richard},
  journal={Communications of the ACM},
  volume={12},
  number={10},
  pages={576--580},
  year={1969},
  publisher={ACM New York, NY, USA}
}

@inproceedings{noble2022more,
  title={More programming than programming: Teaching formal methods in a software engineering programme},
  author={Noble, James and Streader, David and Gariano, Isaac Oscar and Samarakoon, Miniruwani},
  booktitle={NASA Formal Methods Symposium},
  pages={431--450},
  year={2022},
  organization={Springer}
}

@misc{uva_online_judge_2026,
  author = {{UVa Online Judge}},
  title  = {UVa Online Judge},
  howpublished = {\url{https://onlinejudge.org/}},
  year = {n.d.},
  note   = {Accessed: 2026-04-20}
}

@inproceedings{leino2012developing,
  title={Developing verified programs with Dafny},
  author={Leino, K Rustan M},
  booktitle={Proceedings of the 2012 ACM conference on High integrity language technology},
  pages={9--10},
  year={2012}
}

@manual{dafny_official,
  title        = {Dafny Reference Manual},
  author       = {{Dafny Team}},
  organization = {Dafny Software Foundation},
  year         = {2024},
  url          = {https://dafny.org/},
  note         = {Accessed: 2026-04-20}
}

@inproceedings{de2008z3,
  title={Z3: An efficient SMT solver},
  author={De Moura, Leonardo and Bj{\o}rner, Nikolaj},
  booktitle={International conference on Tools and Algorithms for the Construction and Analysis of Systems},
  pages={337--340},
  year={2008},
  organization={Springer}
}

@inproceedings{cook2018formal,
  title={Formal reasoning about the security of amazon web services},
  author={Cook, Byron},
  booktitle={International Conference on Computer Aided Verification},
  pages={38--47},
  year={2018},
  organization={Springer}
}

@inproceedings{wang2023codet5+,
  title={Codet5+: Open code large language models for code understanding and generation},
  author={Wang, Yue and Le, Hung and Gotmare, Akhilesh and Bui, Nghi and Li, Junnan and Hoi, Steven},
  booktitle={Proceedings of the 2023 conference on empirical methods in natural language processing},
  pages={1069--1088},
  year={2023}
}

@inproceedings{le2011boogie,
  title={The boogie verification debugger (tool paper)},
  author={Le Goues, Claire and Leino, K Rustan M and Moskal, Micha{\l}},
  booktitle={International Conference on Software Engineering and Formal Methods},
  pages={407--414},
  year={2011},
  organization={Springer}
}

@article{copet2025cwm,
  title={Cwm: An open-weights llm for research on code generation with world models},
  author={Copet, Jade and Carbonneaux, Quentin and Cohen, Gal and Gehring, Jonas and Kahn, Jacob and Kossen, Jannik and Kreuk, Felix and McMilin, Emily and Meyer, Michel and Wei, Yuxiang and others},
  journal={arXiv preprint arXiv:2510.02387},
  year={2025}
}

@article{kamath2025gemma,
  title={Gemma 3 technical report},
  author={Kamath, Aishwarya and Ferret, Johan and Pathak, Shreya and Vieillard, Nino and Merhej, Ramona and Perrin, Sarah and Matejovicova, Tatiana and Ram{\'e}, Alexandre and Rivi{\`e}re, Morgane and Rouillard, Louis and others},
  journal={arXiv preprint arXiv:2503.19786},
  volume={4},
  year={2025},
  publisher={ArXiv}
}

@article{cao2026qwen3,
  title={Qwen3-coder-next technical report},
  author={Cao, Ruisheng and Chen, Mouxiang and Chen, Jiawei and Cui, Zeyu and Feng, Yunlong and Hui, Binyuan and Jing, Yuheng and Li, Kaixin and Li, Mingze and Lin, Junyang and others},
  journal={arXiv preprint arXiv:2603.00729},
  year={2026}
}

@inproceedings{reynolds2021prompt,
  title={Prompt programming for large language models: Beyond the few-shot paradigm},
  author={Reynolds, Laria and McDonell, Kyle},
  booktitle={Extended abstracts of the 2021 CHI conference on human factors in computing systems},
  pages={1--7},
  year={2021}
}

@inproceedings{tihanyi2025new,
  title={A new era in software security: Towards self-healing software via large language models and formal verification},
  author={Tihanyi, Norbert and Charalambous, Yiannis and Jain, Ridhi and Ferrag, Mohamed Amine and Cordeiro, Lucas C},
  booktitle={2025 IEEE/ACM International Conference on Automation of Software Test (AST)},
  pages={136--147},
  year={2025},
  organization={IEEE}
}

@article{gulwani2017program,
  title={Program synthesis},
  author={Gulwani, Sumit and Polozov, Oleksandr and Singh, Rishabh},
  journal={Foundations and Trends in Programming Languages},
  volume={4},
  number={1-2},
  pages={1--119},
  year={2017},
  publisher={Emerald Publishing Limited}
}

@article{ringer2019qed,
  title={QED at large: A survey of engineering of formally verified software},
  author={Ringer, Talia and Palmskog, Karl and Sergey, Ilya and Milos, Gligoric and Tatlock, Zachary},
  journal={Foundations and Trends in Programming Languages},
  volume={5},
  number={2-3},
  pages={102--281},
  year={2019},
  publisher={Emerald Publishing Limited}
}

@book{jones2021theories,
  title={Theories of programming: the life and works of Tony Hoare},
  author={Jones, Cliff B and Misra, Jayadev},
  year={2021},
  publisher={ACM}
}

@article{yang2023towards,
  title={Towards a correct-by-construction FHE model},
  author={Yang, Zhenkun and Wang, Wen and Casas, Jeremy and Cocchini, Pasquale and Yang, Jin},
  journal={Cryptology ePrint Archive},
  year={2023}
}

@inproceedings{cassez2023formal,
  title={Formal and executable semantics of the ethereum virtual machine in dafny},
  author={Cassez, Franck and Fuller, Joanne and Ghale, Milad K and Pearce, David J and Quiles, Horacio MA},
  booktitle={International Symposium on Formal Methods},
  pages={571--583},
  year={2023},
  organization={Springer}
}

@article{li2022qafny,
  title={Qafny: A quantum-program verifier},
  author={Li, Liyi and Zhu, Mingwei and Cleaveland, Rance and Nicolellis, Alexander and Lee, Yi and Chang, Le and Wu, Xiaodi},
  journal={arXiv preprint arXiv:2211.06411},
  year={2022}
}

@inproceedings{garavel20202020,
  title={The 2020 expert survey on formal methods},
  author={Garavel, Hubert and Ter Beek, Maurice H and Van De Pol, Jaco},
  booktitle={International Conference on Formal Methods for Industrial Critical Systems},
  pages={3--69},
  year={2020},
  organization={Springer}
}

@inproceedings{irfan2022testing,
  title={Testing Dafny (experience paper)},
  author={Irfan, Ahmed and Porncharoenwase, Sorawee and Rakamari{\'c}, Zvonimir and Rungta, Neha and Torlak, Emina},
  booktitle={Proceedings of the 31st ACM SIGSOFT International Symposium on Software Testing and Analysis},
  pages={556--567},
  year={2022}
}

@inproceedings{chakarov2022better,
  title={Better counterexamples for Dafny},
  author={Chakarov, Aleksandar and Fedchin, Aleksandr and Rakamari{\'c}, Zvonimir and Rungta, Neha},
  booktitle={International Conference on Tools and Algorithms for the Construction and Analysis of Systems},
  pages={404--411},
  year={2022},
  organization={Springer}
}

@inproceedings{first2023baldur,
  title={Baldur: Whole-proof generation and repair with large language models},
  author={First, Emily and Rabe, Markus N and Ringer, Talia and Brun, Yuriy},
  booktitle={Proceedings of the 31st ACM Joint European Software Engineering Conference and Symposium on the Foundations of Software Engineering},
  pages={1229--1241},
  year={2023}
}

@article{jiang2022thor,
  title={Thor: Wielding hammers to integrate language models and automated theorem provers},
  author={Jiang, Albert Qiaochu and Li, Wenda and Tworkowski, Szymon and Czechowski, Konrad and Odrzyg{\'o}{\'z}d{\'z}, Tomasz and Mi{\l}o{\'s}, Piotr and Wu, Yuhuai and Jamnik, Mateja},
  journal={Advances in Neural Information Processing Systems},
  volume={35},
  pages={8360--8373},
  year={2022}
}

@article{wu2022autoformalization,
  title={Autoformalization with large language models},
  author={Wu, Yuhuai and Jiang, Albert Qiaochu and Li, Wenda and Rabe, Markus and Staats, Charles and Jamnik, Mateja and Szegedy, Christian},
  journal={Advances in neural information processing systems},
  volume={35},
  pages={32353--32368},
  year={2022}
}

@inproceedings{madaan2022language,
  title={Language models of code are few-shot commonsense learners},
  author={Madaan, Aman and Zhou, Shuyan and Alon, Uri and Yang, Yiming and Neubig, Graham},
  booktitle={Proceedings of the 2022 Conference on Empirical Methods in Natural Language Processing},
  pages={1384--1403},
  year={2022}
}

@article{frieder2023mathematical,
  title={Mathematical capabilities of chatgpt},
  author={Frieder, Simon and Pinchetti, Luca and Chevalier, Chevalier and Griffiths, Ryan-Rhys and Salvatori, Tommaso and Lukasiewicz, Thomas and Petersen, Philipp and Berner, Julius},
  journal={Advances in neural information processing systems},
  volume={36},
  pages={27699--27744},
  year={2023}
}

@inproceedings{narkawicz2017minerva,
  title={The MINERVA software development process},
  author={Narkawicz, Anthony and Munoz, C{\'e}sar A and Dutle, Aaron M},
  booktitle={NASA Formal Methods Symposium (NFM) 2017},
  number={NF1676L-26800},
  year={2017}
}

@techreport{lewis2001model,
  title={Model-based verification: Analysis guidelines},
  author={Lewis, Grace A and Comella-Dorda, Santiago and Gluch, David P and Hudak, John and Weinstock, Charles},
  year={2001}
}

@inproceedings{nashid2023retrieval,
  title={Retrieval-based prompt selection for code-related few-shot learning},
  author={Nashid, Noor and Sintaha, Mifta and Mesbah, Ali},
  booktitle={2023 IEEE/ACM 45th International Conference on Software Engineering (ICSE)},
  pages={2450--2462},
  year={2023},
  organization={IEEE}
}

@inproceedings{tufano2023automating,
  title={Automating code-related tasks through transformers: The impact of pre-training},
  author={Tufano, Rosalia and Pascarella, Luca and Bavota, Gabriele},
  booktitle={2023 IEEE/ACM 45th International Conference on Software Engineering (ICSE)},
  pages={2425--2437},
  year={2023},
  organization={IEEE}
}

@article{grattafiori2024llama,
  title={The llama 3 herd of models},
  author={Grattafiori, Aaron and Dubey, Abhimanyu and Jauhri, Abhinav and Pandey, Abhinav and Kadian, Abhishek and Al-Dahle, Ahmad and Letman, Aiesha and Mathur, Akhil and Schelten, Alan and Vaughan, Alex and others},
  journal={arXiv preprint arXiv:2407.21783},
  year={2024}
}

@article{yang2025qwen3,
  title={Qwen3 technical report},
  author={Yang, An and Li, Anfeng and Yang, Baosong and Zhang, Beichen and Hui, Binyuan and Zheng, Bo and Yu, Bowen and Gao, Chang and Huang, Chengen and Lv, Chenxu and others},
  journal={arXiv preprint arXiv:2505.09388},
  year={2025}
}

@article{sellergren2025medgemma,
  title={Medgemma technical report},
  author={Sellergren, Andrew and Kazemzadeh, Sahar and Jaroensri, Tiam and Kiraly, Atilla and Traverse, Madeleine and Kohlberger, Timo and Xu, Shawn and Jamil, Fayaz and Hughes, C{\'\i}an and Lau, Charles and others},
  journal={arXiv preprint arXiv:2507.05201},
  year={2025}
}

@inproceedings{sun2024clover,
  title={Clover: Clo sed-loop ver ifiable code generation},
  author={Sun, Chuyue and Sheng, Ying and Padon, Oded and Barrett, Clark},
  booktitle={International Symposium on AI Verification},
  pages={134--155},
  year={2024},
  organization={Springer}
}

@article{austin2021program,
  title={Program synthesis with large language models},
  author={Austin, Jacob and Odena, Augustus and Nye, Maxwell and Bosma, Maarten and Michalewski, Henryk and Dohan, David and Jiang, Ellen and Cai, Carrie and Terry, Michael and Le, Quoc and others},
  journal={arXiv preprint arXiv:2108.07732},
  year={2021}
}

@article{baksys2025atlas,
  title={ATLAS: Automated Toolkit for Large-Scale Verified Code Synthesis},
  author={Baksys, Mantas and Zetzsche, Stefan and Bouissou, Olivier and Delmas, Remi and Kong, Soonho and Holden, Sean B},
  journal={arXiv preprint arXiv:2512.10173},
  year={2025}
}

@article{loughridge2024dafnybench,
  title={Dafnybench: A benchmark for formal software verification},
  author={Loughridge, Chloe and Sun, Qinyi and Ahrenbach, Seth and Cassano, Federico and Sun, Chuyue and Sheng, Ying and Mudide, Anish and Misu, Md Rakib Hossain and Amin, Nada and Tegmark, Max},
  journal={arXiv preprint arXiv:2406.08467},
  year={2024}
}

@inproceedings{ma2025specgen,
  title={Specgen: Automated generation of formal program specifications via large language models},
  author={Ma, Lezhi and Liu, Shangqing and Li, Yi and Xie, Xiaofei and Bu, Lei},
  booktitle={2025 IEEE/ACM 47th International Conference on Software Engineering (ICSE)},
  pages={16--28},
  year={2025},
  organization={IEEE}
}

@inproceedings{siegmund2015views,
  title={Views on internal and external validity in empirical software engineering},
  author={Siegmund, Janet and Siegmund, Norbert and Apel, Sven},
  booktitle={2015 IEEE/ACM 37th IEEE International Conference on Software Engineering},
  volume={1},
  pages={9--19},
  year={2015},
  organization={IEEE}
}

@inproceedings{feldt2010validity,
  title={Validity threats in empirical software engineering research-an initial survey.},
  author={Feldt, Robert and Magazinius, Ana},
  booktitle={Seke},
  pages={374--379},
  year={2010}
}

@article{wangtoward,
  title={Toward Automated, Contamination-free Dafny Benchmark Generation},
  author={Wang, Changjie and Scazzariello, Mariano and Kosti{\'c}, Dejan and Chiesa, Marco}
}

@article{banerjee2026dafnypro,
  title={DafnyPro: LLM-Assisted Automated Verification for Dafny Programs},
  author={Banerjee, Debangshu and Bouissou, Olivier and Zetzsche, Stefan},
  journal={arXiv preprint arXiv:2601.05385},
  year={2026}
}

@article{white2023prompt,
  title={A prompt pattern catalog to enhance prompt engineering with chatgpt},
  author={White, Jules and Fu, Quchen and Hays, Sam and Sandborn, Michael and Olea, Carlos and Gilbert, Henry and Elnashar, Ashraf and Spencer-Smith, Jesse and Schmidt, Douglas C},
  journal={arXiv preprint arXiv:2302.11382},
  year={2023}
}

@article{giray2023prompt,
  title={Prompt engineering with ChatGPT: a guide for academic writers},
  author={Giray, Louie},
  journal={Annals of biomedical engineering},
  volume={51},
  number={12},
  pages={2629--2633},
  year={2023},
  publisher={Springer}
}

@article{mesbah2011invariant,
  title={Invariant-based automatic testing of modern web applications},
  author={Mesbah, Ali and Van Deursen, Arie and Roest, Danny},
  journal={IEEE Transactions on Software Engineering},
  volume={38},
  number={1},
  pages={35--53},
  year={2011},
  publisher={IEEE}
}

@online{MicrosoftCopilot,
  title        = {Microsoft Copilot: Your AI Companion},
  url          = {https://copilot.microsoft.com/},
  note         = {Accessed: 2026-04-21},
}

@misc{cursor2024,
  author = {{Anysphere, Inc.}},
  title = {Cursor: The {AI} Code Editor},
  year = {2024},
  url = {https://cursor.com/},
  note = {Accessed: 2026-04-21}
}

@misc{amazonq2024,
  author = {{Amazon Web Services, Inc.}},
  title = {Amazon {Q} Developer: {AI} coding companion},
  year = {2024},
  url = {https://aws.amazon.com/q/developer/},
  note = {Accessed: 2026-04-21}
}

@article{ray2025review,
  title={A review on vibe coding: Fundamentals, state-of-the-art, challenges and future directions},
  author={Ray, Partha Pratim},
  journal={Authorea Preprints},
  year={2025},
  publisher={Authorea}
}

@article{ji2023survey,
  title={Survey of hallucination in natural language generation},
  author={Ji, Ziwei and Lee, Nayeon and Frieske, Rita and Yu, Tiezheng and Su, Dan and Xu, Yan and Ishii, Etsuko and Bang, Ye Jin and Madotto, Andrea and Fung, Pascale},
  journal={ACM computing surveys},
  volume={55},
  number={12},
  pages={1--38},
  year={2023},
  publisher={ACM New York, NY}
}

@misc{uva_11934,
  author       = {{UVa Online Judge}},
  title        = {Problem 11934: {M}agic {F}ormula},
  howpublished = {\url{https://onlinejudge.org/external/119/11934.pdf}},
  year         = {2010},
  note         = {Accessed: 2026-04-21}
}

@article{ter2024role,
  title={The role of formal methods in computer science education},
  author={ter Beek, Maurice and Broy, Manfred and Dongol, Brijesh},
  journal={ACM Inroads},
  volume={15},
  number={4},
  pages={58--66},
  year={2024},
  publisher={ACM New York, NY, USA}
}

@article{paul2023formal,
  title={Formal verification of safety-critical aerospace systems},
  author={Paul, Saswata and Cruz, Elkin and Dutta, Airin and Bhaumik, Ankita and Blasch, Erik and Agha, Gul and Patterson, Stacy and Kopsaftopoulos, Fotis and Varela, Carlos},
  journal={IEEE Aerospace and Electronic Systems Magazine},
  volume={38},
  number={5},
  pages={72--88},
  year={2023},
  publisher={IEEE}
}

@inproceedings{dipu2024formalfuzzer,
  title={Formalfuzzer: Formal verification assisted fuzz testing for soc vulnerability detection},
  author={Dipu, Nusrat Farzana and Hossain, Muhammad Monir and Azar, Kimia Zamiri and Farahmandi, Farimah and Tehranipoor, Mark},
  booktitle={2024 29th Asia and South Pacific Design Automation Conference (ASP-DAC)},
  pages={355--361},
  year={2024},
  organization={IEEE}
}

@online{GoogleVibeCoding2024,
  author = {{Google}},
  title = {Vibe Coding in Google AI Studio: Building Apps with Natural Language},
  year = {2024},
  url = {https://aistudio.google.com/vibe-code},
  note = {Accessed: Apr. 22, 2026}
}

@misc{MSRDafnyProject,
  author = {{Microsoft Research}},
  title  = {Dafny: A Language and Program Verifier for Functional Correctness},
  year   = {2024},
  url    = {https://www.microsoft.com/en-us/research/project/dafny-a-language-and-program-verifier-for-functional-correctness/},
  note   = {Official Project Page}
}

@article{AmazonDafny2023,
  author  = {Aleksandr Fedchin and Tyler Dean and Jeffrey S. Foster and others},
  title   = {A Toolkit for Automated Testing of {Dafny}},
  journal = {Amazon Science},
  year    = {2023},
  url     = {https://www.amazon.science/publications/a-toolkit-for-automated-testing-of-dafny}
}

@article{manik2026gemma,
  title={Gemma 4, Phi-4, and Qwen3: Accuracy-Efficiency Tradeoffs in Dense and MoE Reasoning Language Models},
  author={Manik, Md Motaleb Hossen and Wang, Ge},
  journal={arXiv preprint arXiv:2604.07035},
  year={2026}
}

@inproceedings{czajka2018concrete,
  title={Concrete semantics with Coq and CoqHammer},
  author={Czajka, {\L}ukasz and Ekici, Burak and Kaliszyk, Cezary},
  booktitle={International Conference on Intelligent Computer Mathematics},
  pages={53--59},
  year={2018},
  organization={Springer}
}

@manual{DafnyRefManual,
  title        = {Dafny Reference Manual},
  author       = {{The Dafny Project}},
  organization = {Amazon Web Services},
  year         = {2024},
  url          = {https://dafny.org/latest/DafnyRef/DafnyRef},
  note         = {Accessed: Apr. 22, 2026}
}

@article{misu2024towards,
  title={Towards ai-assisted synthesis of verified dafny methods},
  author={Misu, Md Rakib Hossain and Lopes, Cristina V and Ma, Iris and Noble, James},
  journal={Proceedings of the ACM on Software Engineering},
  volume={1},
  number={FSE},
  pages={812--835},
  year={2024},
  publisher={ACM New York, NY, USA}
}

@misc{gpt_oss,
  author = {{OpenAI}},
  title  = {Introducing {GPT-OSS}: Open-Weight Reasoning Models},
  year   = {2025},
  url    = {https://openai.com/index/introducing-gpt-oss/},
  note   = {Accessed: Apr. 22, 2026}
}

@misc{qwen_series,
  author = {{Alibaba Qwen Team}},
  title  = {Qwen3.5 and Qwen3.6-MoE: Advancing Open-Weight Foundation Models},
  year   = {2026},
  url    = {https://github.com/QwenLM/Qwen},
  note   = {Accessed: Apr. 22, 2026}
}

@misc{qwen3_coder,
  author = {{Alibaba Qwen Team}},
  title  = {Qwen3-Coder-30B: Specialized Models for Agentic Code Intelligence},
  year   = {2025},
  url    = {https://huggingface.co/Qwen},
  note   = {Accessed: Apr. 22, 2026}
}

@misc{gemma4,
  author = {{Google DeepMind}},
  title  = {Gemma 4: Next-Generation Open Multimodal Models},
  year   = {2026},
  url    = {https://ai.google.dev/gemma},
  note   = {Accessed: Apr. 22, 2026}
}

@misc{codestral_v01,
  author = {{Mistral AI}},
  title  = {Codestral-22B-v0.1: An Open-Weight Model for Professional Coders},
  year   = {2024},
  url    = {https://mistral.ai/news/codestral/},
  note   = {Accessed: Apr. 22, 2026}
}

@misc{DafnyNuGet,
  author       = {{The Dafny Project}},
  title        = {Dafny {NuGet} Package: The {Dafny} Compiler and Verifier},
  year         = {2024},
  url          = {https://www.nuget.org/packages/Dafny/},
  note         = {Version 4.x.x, Accessed: Apr. 22, 2026},
  howpublished = {NuGet Package Manager}
}

@article{rushby1995model,
  title={Model checking and other ways of automating formal methods},
  author={Rushby, JM},
  journal={Position paper for panel on model checking for concurrent programs, Software Quality Week, San Francisco},
  year={1995}
}

@misc{github_dafny_2026,
  author = {{GitHub Community}},
  title = {Dafny Repositories Search Results},
  year = {2026},
  url = {https://github.com/search?q=dafny&type=repositories},
  note = {Accessed: 2026-04-23}
}

\end{document}